\documentclass[manuscript, screen]{acmart}

\usepackage{amsmath,amsfonts}
\usepackage{algorithmic}
\usepackage{graphicx}
\usepackage{textcomp}
\usepackage{xcolor}

\usepackage{multirow}
\usepackage{subfigure}
\usepackage{adjustbox}
\usepackage{wrapfig}
\usepackage{xspace}
\usepackage{pifont}
\usepackage{soul}
\usepackage{stfloats}
\usepackage{float}
\usepackage{colortbl}
\usepackage{booktabs}
\usepackage{caption}
\usepackage{bbding}
\newcommand{\cmark}{\ding{51}}%
\newcommand{\xmark}{\ding{55}}%
\AtBeginDocument{%
  }

\setcopyright{acmlicensed}
\copyrightyear{2018}
\acmYear{2018}
\acmDOI{XXXXXXX.XXXXXXX}
\acmConference[Conference acronym 'XX]{Make sure to enter the correct
  conference title from your rights confirmation email}{June 03--05,
  2018}{Woodstock, NY}
\acmISBN{978-1-4503-XXXX-X/2018/06}

\begin{document}

\title{Does Multimodality Improve Recommender Systems as Expected? A Critical Analysis and Future Directions}


\author{Hongyu Zhou}
\affiliation{%
  \institution{Nanyang Technological University}
  \country{Singapore}}
\email{hongyu006@e.ntu.edu.sg}

\author{Yinan Zhang}
\affiliation{%
  \institution{Nanyang Technological University}
  \country{Singapore}}
\email{yinan.zhang@ntu.edu.sg}

\author{Aixin Sun}
\affiliation{%
  \institution{Nanyang Technological University}
  \country{Singapore}}
\email{axsun@ntu.edu.sg}

\author{Zhiqi Shen}
\affiliation{%
  \institution{Nanyang Technological University}
  \country{Singapore}}
\email{zqshen@ntu.edu.sg}


\begin{abstract}
  Multimodal recommendation systems are increasingly popular for their potential to improve performance by integrating diverse data types. However, the actual benefits of this integration remain unclear, raising questions about when and how it truly enhances recommendations. In this paper, we propose a structured evaluation framework to systematically assess multimodal recommendations across four dimensions: Comparative Efficiency, Recommendation Tasks, Recommendation Stages, and Multimodal Data Integration. We benchmark a set of reproducible multimodal models against strong traditional baselines and evaluate their performance on different platforms. Our findings show that multimodal data is particularly beneficial in sparse interaction scenarios and during the recall stage of recommendation pipelines. We also observe that the importance of each modality is task-specific, where text features are more useful in e-commerce and visual features are more effective in short-video recommendations. Additionally, we explore different integration strategies and model sizes, finding that Ensemble-Based Learning outperforms Fusion-Based Learning, and that larger models do not necessarily deliver better results. To deepen our understanding, we include case studies and review findings from other recommendation domains. Our work provides practical insights for building efficient and effective multimodal recommendation systems, emphasizing the need for thoughtful modality selection, integration strategies, and model design.
\end{abstract}



\received{20 February 2007}
\received[revised]{12 March 2009}
\received[accepted]{5 June 2009}

\maketitle

\section{Introduction}
Recent years have witnessed the widespread application of multimodal information across various domains such as healthcare~\cite{soenksen2022integrated} and autonomous driving~\cite{prakash2021multi}, offering rich, complementary perspectives that enhance machine understanding and decision-making processes~\cite{baltruvsaitis2018multimodal}. Integrating data from different modalities is particularly promising in recommendation, which benefits from the rich diversity of content features and helps alleviate the data sparsity and cold start issues~\cite{zhang2019deep}. Traditional recommendation systems, often based solely on user-item interaction, may overlook the subtleties of user preferences that modality data can capture~\cite{deldjoo2020recommender}. Modality data not only reflects the relations between users and items but also encodes item characteristics and user intent, which is particularly valuable in scenarios with sparse interaction data. Additionally, different modalities provide complementary perspectives on item attributes, thereby enhancing the richness of the information~\cite{wang2022deep}. For example, in e-commerce, analyzing images and text together often provides deeper insights into customer preferences than either modality alone, thus enriching the recommendation capabilities where limited data is available~\cite{liu2023multimodal2}. 

To harness this potential, existing multimodal recommendation models typically integrate content features extracted from various modalities with collaborative filtering signals derived from user-item interactions~\cite{he2016vbpr, wei2020graph}. 
However, despite increasing research interest in multimodal recommendations, there remains a significant gap in systematic evaluations of their efficacy~\cite{deldjoo2020recommender}. Notably, some recent studies have even suggested that unimodal models can outperform multimodal ones in specific scenarios~\cite{zhou2023comprehensive}, raising fundamental questions about the value and application of multimodal data in recommendation systems. This observation motivates several key research questions:
\begin{enumerate}
    \item Does multimodal information consistently improve recommendation performance?
    \item What is the relative contribution of each modality?
    \item Are multimodal approaches especially beneficial for users or items with sparse interaction histories?
    \item How does the effectiveness of multimodal data differ across various recommendation tasks, such as e-commerce and short video platforms?
    \item Which stages of the recommendation pipeline (e.g., recall vs. reranking) gain the most from multimodal input?
    \item What types of architectures are most effective for integrating multimodal data?
\end{enumerate}


To answer these questions, we propose a structured framework for evaluating the effectiveness of multimodal recommendation systems. This framework is organized around four dimensions: Comparative Efficiency, Recommendation Tasks, Recommendation Stages, and Multimodal Data Integration. 
We begin by reviewing a set of representative multimodal recommendation models that are publicly available and reproducible. These models are then benchmarked against strong traditional baselines that rely solely on user-item interactions to investigate whether modality data is useful. To evaluate the incremental value of multimodal inputs, we analyze the contribution of each modality to overall performance. We also examine how performance varies under different sparsity degrees from user and item sides, aiming to assess whether multimodal data offers greater benefits in sparse data scenarios.

Our evaluation spans a variety of recommendation tasks and stages to provide a comprehensive view of model effectiveness. To this end, we compare the performance of all code reproducible models under the same evaluation settings across various platforms and sources, including Amazon, Taobao, and the short video platform DY, to explore the utility of modality under different recommendation scenarios. To identify at which stage modality is beneficial, we compare multimodal models against the best traditional recommenders at the recall and reranking stages of recommendation. Furthermore, we investigate different strategies for integrating multimodal data, including early fusion and late ensemble methods, to determine which approaches are most effective in improving the accuracy of the recommendation. 

Beyond empirical evaluation, we expand our analysis to additional domains such as news, food, music, and POI recommendation, highlighting how the importance of different modalities varies across tasks. We also examine the relationship between model size and effectiveness, revealing that larger models do not necessarily lead to better performance, and that efficient modality integration is often more critical than model scale. To gain deeper insights into real-world behavior, we conduct case studies comparing traditional and multimodal methods, illustrating when and why multimodal signals are helpful. Through this in-depth analysis, we aim to provide practical guidance on how to optimally leverage multimodal information in recommendation systems.

In summary, the main contribution is outlined below. 
\begin{itemize}
    \item We evaluate the reproducibility of existing research in multimodal recommendation systems and compare their efficiency against traditional recommenders. Our findings indicate that traditional methods, which rely solely on user-item interaction history, perform comparably or even better than some multimodal approaches, highlighting the limited utilization of additional modality information. However, we observe that multimodal recommendation methods tend to outperform traditional ones more noticeably in cases involving users or items with limited interaction data.
    \item Different types of modal data exhibit varying levels of importance across different recommendation tasks. In e-commerce recommendations, text data is usually more beneficial than images, whereas in short-video recommendation tasks, images tend to be more advantageous. Our review of existing studies in other domains reveals similar patterns, reinforcing the idea that the effectiveness of each modality depends on the specific recommendation scenario.
    \item Multimodal data tends to be more beneficial in the recall stage than in the reranking stage of recommendations. As the number of items ($N$) in Top-$N$ recommendations increases, multimodal recommendation systems progressively exhibit superior performance improvements compared to traditional methods.
    \item We explore different methods of learning multimodal representation, particularly focusing on Fusion-Based and Ensemble-Based Learning. The former merges modal features early to form a unified prediction, whereas the latter makes modality-specific predictions and integrates these predictions later. Our findings demonstrate that Ensemble-Based often outperforms Fusion-Based Learning, indicating that managing modal features independently before integration enhances the effectiveness of multimodal recommendation systems.

    
\end{itemize}

\begin{figure}[ht]
    \centering
    \includegraphics[width=\linewidth]{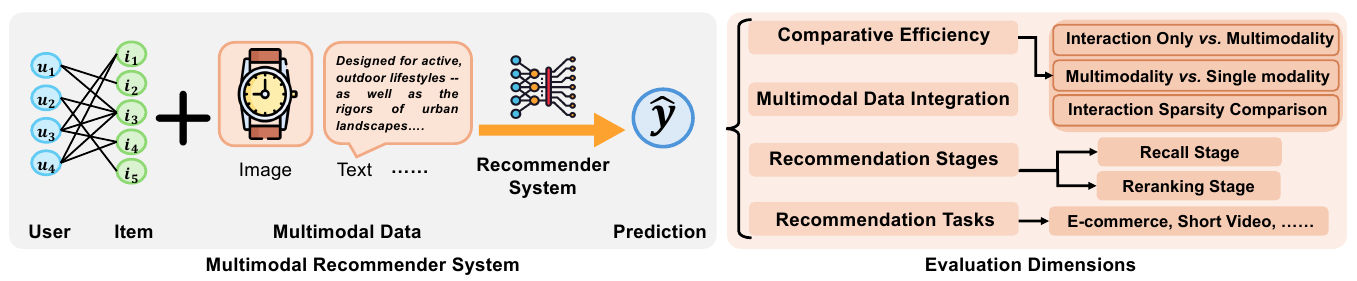}
    \caption{Framework for evaluating the impact of multimodal data on recommendation systems across four key dimensions: Comparative Efficiency, Multimodal Data Integration, Recommendation Stages, and Recommendation Tasks.}
    \label{framework}
\end{figure}

\section{Research Method}
As illustrated in Figure~\ref{framework}, we propose a framework for comprehensive evaluation of the effectiveness of multimodal data in recommendation systems, which consists of four key dimensions: \textit{Comparative Efficiency}, \textit{Recommendation Tasks}, \textit{Recommendation Stages}, and \textit{Multimodal Data Integration}.

In the \textit{Comparative Efficiency} aspect, our framework evaluates how multimodal information enhances recommendation system accuracy. Firstly, we compare the performance of multimodal recommendation systems against traditional systems that solely rely on user-item interactions. Secondly, we evaluate the recommendation accuracy when a specific modality is excluded from the multimodal system. This approach enables us to isolate and analyze the individual contributions of each modality to the overall effectiveness of the recommendation system. Lastly, we compare the performance between traditional models with multimodal models under different interaction sparsity degrees, which analyzes the influence of interaction sparsity on the performance of recommendation systems. 

To more comprehensively analyze the impact of modal data on recommendation systems, we propose to explore how multimodality influences model performance across various \textit{Recommendation Tasks}, such as e-commerce and short video recommendations. Recognizing that different modalities may have varying effects depending on the recommendation context, this evaluation allows us to tailor and optimize the use of multimodality, enhancing the relevance and accuracy of recommendations specific to each task.

We also propose to evaluate how multimodality influences various \textit{Recommendation Stages}, such as recall and reranking. By systematically assessing these effects, we aim to provide insights into how to optimally leverage multimodality to enhance both the breadth of initial item retrieval and the accuracy of subsequent ranking.

Finally, we explored how integrating multimodal data can enhance the utility of recommendation systems. We analyzed how state-of-the-art models learn multimodal representations, categorizing them into Fusion-Based and Ensemble-Based Learning, and assessed their respective performances. We believe that this can provide critical insight and guidance for improving performance in the design of future \textit{Multimodal Data Integration}.

\subsection{Reproducible Paper Collection}
\label{sec:PaperCollection}
To ensure our analysis extends beyond individual case studies and results that can be validated, we conducted a comprehensive review of conference proceedings and journal publications to identify pertinent works on multimodal recommendations. 
This thorough review helps establish a solid foundation for our study, enabling us to leverage reliable and reproducible research in the field of multimodal recommendation systems.
Our analysis focused on papers published between 2019 and 2024 in top-tier conferences and journals\footnote{This list is not intended to be comprehensive; however, we believe it captures the most significant contributions in this field.}, including TMM, ACM MM, SIGIR, TheWebConf (WWW), TKDE, CIKM, TOIS, and AAAI. We considered a paper relevant if it introduced a new technique and tackled issues related to multimodal recommendations. Studies that were exclusively focused on single-modality were excluded. This approach resulted in a compiled dataset of 41 pertinent papers.

Then, we attempted to reproduce the results reported in the selected papers using the original artifacts provided by the authors, including source code and datasets. Ideally, results should be reproducible based solely on the technical descriptions in the papers. However, in practice, subtle implementation details and evaluation procedures, such as data splitting and hyperparameter settings, can critically influence experimental outcomes. A paper was considered \textit{Reproducible} if it met the following criteria: the official source code is available and functions correctly. Regarding the datasets, research that relies exclusively on non-public, company-owned data or on data collected from the web but not publicly shared was excluded from consideration as reproducible. The datasets will be considered as reproducible if they are publicly accessible as used in the original study or the authors provided raw data with the preprocessing code. Based on these criteria, we categorized the research works into \textit{Code Reproducible}, \textit{Dataset Available}, and \textit{Non-reproducible}.

The fractions of reproducible papers are displayed in Table~\ref{tab:reproducible_works}. Overall, approximately 29.3\% of the papers are \textit{Reproducible}, with both \textit{Code Reproducible} and \textit{Dataset Available}, highlighting the prevalent issues of limited reproducibility we have identified. For these papers, we will strictly follow the provided instructions like the optimal hyperparameters or search space to replicate the reported performance results. Additionally, for works where the code is reproducible but the dataset involved in their original experiments is not accessible, we will apply these algorithms on publicly available datasets to enable further performance comparisons.

\begin{table}[]
\centering
\caption{\textit{Reproducible} research on multimodal recommender systems published in top-tier conferences or journals from 2019 to 2024. `\textit{Code Reproducible}' means the official source code is available and functions correctly. `\textit{Dataset Available}' means the datasets used in the original study are accessible. A work that is both `\textit{Code Reproducible}' and `\textit{Dataset Available}' will be considered `\textit{Reproducible}'; otherwise, it will be classified as `\textit{Non-reproducible}'.}
\label{tab:reproducible_works}
\resizebox{0.6\columnwidth}{!}{
\begin{tabular}{ llll}
\toprule
Publication & \multicolumn{1}{|c}{\begin{tabular}[c]{@{}c@{}}\textit{Code}\\ \textit{Reproducible}\end{tabular}} & \multicolumn{1}{|c|}{\begin{tabular}[c]{@{}c@{}}\textit{Dataset}\\ \textit{Available}\end{tabular} }& \textit{Reproducible} \\ 
\midrule
TMM        & \multicolumn{1}{|l}{ \begin{tabular}[l]{@{}l@{}}\cite{chen2020learning}, \cite{wang2021dualgnn}, \cite{wei2021hierarchical},\\ \cite{tao2022self}, \cite{liu2023multimodal}\end{tabular}}
& \multicolumn{1}{|l|}{\cite{liu2023multimodal}, \cite{zhou2024disentangled}, \cite{guo2024space}  } & \cite{liu2023multimodal}\\
ACM MM     & \multicolumn{1}{|l}{\begin{tabular}[l]{@{}l@{}} \cite{wei2019mmgcn}, \cite{wei2020graph}, \cite{zhang2021mining}, \cite{du2022invariant}, \\ \cite{chen2022breaking}, \cite{yu2023multi}, \cite{liu2022elimrec}, \cite{zhou2023tale},\\ \cite{jiang2024diffmm}\end{tabular}}
& \multicolumn{1}{|l}{\begin{tabular}[l]{@{}l@{}} \cite{zhang2021mining}, \cite{mu2022learning}, \cite{zhang2024modality}, \\ \cite{zhou2023tale}, \cite{yu2023multi}, \cite{yang2024multimodal},\\ \cite{jiang2024diffmm} \end{tabular}}
& \multicolumn{1}{|l}{\begin{tabular}[l]{@{}l@{}} \cite{zhang2021mining}, \cite{zhou2023tale}, \cite{yu2023multi},\\ \cite{jiang2024diffmm} \end{tabular}}\\
SIGIR      & \multicolumn{1}{|l}{\cite{wei2023lightgt}, \cite{li2023multimodal}}& \multicolumn{1}{|l|}{\cite{chen2019personalized}, \cite{zhang2023mining}, \cite{zhou2023attention}, \cite{li2023multimodal} } & \cite{li2023multimodal} \\
WWW        & \multicolumn{1}{|l}{\cite{han2022modality}, \cite{wei2023multi}, \cite{zhou2023bootstrap}}& \multicolumn{1}{|l|}{\cite{zhou2023bootstrap} } & \cite{zhou2023bootstrap} \\
Others & \multicolumn{1}{|l}{ \begin{tabular}[l]{@{}l@{}}\cite{zhang2022latent}, \cite{liu2023dynamic}, \cite{li2024multimodal},\\ \cite{guo2024lgmrec}, \cite{xu2025mentor}, \cite{lin2024gume}\end{tabular}}& \multicolumn{1}{|l}{\begin{tabular}[l]{@{}l@{}}\cite{zhang2022latent}, \cite{kim2022mario}, \cite{li2024multimodal},\\ \cite{guo2024lgmrec}, \cite{xu2025mentor}, \cite{lin2024gume} \end{tabular}} & \multicolumn{1}{|l}{\begin{tabular}[l]{@{}l@{}}\cite{zhang2022latent}, \cite{li2024multimodal}, \cite{guo2024lgmrec},\\ \cite{xu2025mentor}, \cite{lin2024gume}\end{tabular}} \\
\midrule
Total. ratio      &  &  & 12/41 (29.3\%) \\
\bottomrule
\multicolumn{4}{p{8.5cm}}{\textit{Non-reproducible}: TMM: \cite{chen2020learning}, \cite{yi2021multi}, \cite{wang2021dualgnn}, \cite{cai2021heterogeneous}, \cite{wei2021hierarchical}, \cite{cai2022heterogeneous}, \cite{tao2022self}, \cite{zhou2024disentangled}, \cite{guo2024space}; ACM MM: \cite{wei2019mmgcn}, \cite{wei2020graph}, \cite{du2022invariant}, \cite{chen2022breaking}, \cite{zhang2024modality}, \cite{liu2022elimrec}, \cite{mu2022learning}, \cite{huang2023pareto}, \cite{yang2024multimodal}, \cite{yang2023modal}; SIGIR: \cite{chen2019personalized}, \cite{yi2022multi}, \cite{wei2023lightgt}, \cite{zhang2023mining}, \cite{zhou2023attention}; WWW: \cite{liu2019user}, \cite{han2022modality}, \cite{wei2023multi}; Other: \cite{liu2023dynamic}, \cite{kim2022mario}}
\end{tabular}
}
\end{table}

Our study explores a range of multimodal recommendation models that stand out due to their \textit{Code Reproducibility} and \textit{Dataset Available}. These models leverage all graph-based techniques and multimodal fusion to enhance item and user representations accurately. 

Some models build supplementary graphs to enhance user and item embeddings that capture more relations than historical interactions. Specifically:
\begin{itemize}
    \item \textbf{LATTICE}\footnote{https://github.com/CRIPAC-DIG/LATTICE}~\cite{zhang2021mining} constructs a homogeneous graph within a modality-aware structure learning layer, which leverages modality features to identify item structures and employs graph convolutions to refine representations by exploring item affinities. 
    \item \textbf{MICRO}\footnote{https://github.com/CRIPAC-DIG/MICRO}~\cite{zhang2022latent} introduces a contrastive framework to enhance multimodal fusion, aligning modality-aware representations with fused multimodal representations using an attention mechanism.
    \item \textbf{FREEDOM}\footnote{https://github.com/enoche/FREEDOM}~\cite{zhou2023tale} finds that freezing the item-item graph prior to training improves the learning of item semantic relations. Additionally, it uses degree-sensitive edge pruning to denoise the graph.
    \item \textbf{MGCE}\footnote{\label{MGCE}https://github.com/hfutmars/MGCE} \cite{li2024multimodal} utilizes causal theory to extract distinct multimodal interest and conformity embeddings from the structural and feature-based aspects of graph.
    \item \textbf{LGMRec}\footnote{https://github.com/georgeguo-cn/LGMRec?tab=readme-ov-file}~\cite{guo2024lgmrec} effectively addresses both local and global user interests. It utilizes a local graph embedding module to capture collaborative and modality-related embeddings based on immediate relationships, while a global hypergraph module further extends these insights by incorporating global dependencies. 
    \item \textbf{GUME}\footnote{https://github.com/NanGongNingYi/GUME}~\cite{lin2024gume} utilizes the learned item-item semantic relations to enhance the user-item graph. It leverages the item modality semantic graph to extract explicit interaction features and learned the extended interest features from the enhanced user-item graph. It reaggregates the coarse-grained and fine-grained attributes to form enhanced explicit-interaction features. A user-modality enhancement module and an alignment module are designed to improve the performance.
\end{itemize}

Several research efforts are dedicated to noise reduction in multimodal recommendation systems:
\begin{itemize}
    \item \textbf{BM3}\footnote{https://github.com/enoche/BM3}~\cite{zhou2023bootstrap} introduces self-supervised learning into multimodal recommendation to boost computational efficiency and reduce noise. It employs dropout techniques to generate diverse representations and optimizes multi-modal objectives, thereby enhancing performance with lower computational demand. 
    \item \textbf{MGCN}\footnote{https://github.com/demonph10/MGCN}~\cite{yu2023multi} enhances modality feature learning by filtering out noise using item interaction information, then enriches these purified features across distinct user-item and item-item views. This method improves feature clarity and effectiveness within the model. 
    \item \textbf{MGCL}\footnote{https://github.com/hfutmars/MGCL}~\cite{liu2023multimodal} deploys three parallel Graph Neural Networks that independently generate user and item representations based on collaborative signals, visual preferences, and textual preferences. To further reduce noise, MGCL incorporates robust self-supervised contrastive learning into its optimization process. 
    \item \textbf{MCLN}\footnote{https://github.com/hfutmars/MCLN}~\cite{li2023multimodal} proposes a counterfactual learning module to distinguish between the preference distributions of modality from user-interacted and uninteracted items, helping to identify and eliminate irrelevant components.

\end{itemize}

Recent research has introduced self-supervised learning techniques to enhance recommendation systems. For example, BM3 generates contrastive views and designs a contrastive loss function to jointly optimize three objectives. However, simplistic random augmentation or heuristic cross-view strategies often introduce irrelevant noise or loss the valuable interaction information.
\begin{itemize}
    \item \textbf{DiffMM}\footnote{https://github.com/HKUDS/DiffMM}~\cite{jiang2024diffmm} proposes a Multi-Modal Graph Diffusion Model to align multi-modal context with user-item interaction modeling, constructs user-item interaction graphs that incorporate modality information, progressively corrupts the original graph's interactions, and employs iterative learning to recover the original structure through a probabilistic diffusion process.
    \item \textbf{MENTOR}\footnote{https://github.com/Jinfeng-Xu/MENTOR}~\cite{xu2025mentor} designs a multilevel self-supervised learning task that aligns modalities under the direct or indirect guidance of id embeddings and aligns the fused modality with id embedding. The new designed task could encourage align cross-modality and retrain the interaction information.
\end{itemize}

Our study also includes the following multimodal recommendation models that are \textit{Code Reproducible} only:
\begin{itemize}
    \item \textbf{VBPR}\footnote{https://github.com/arogers1/VBPR}~\cite{he2016vbpr} is the first model that considers introducing visual features into the recommendation system by concatenating visual embeddings with id embeddings as the item representation. 
    \item \textbf{MMGCN}\footnote{https://github.com/weiyinwei/MMGCN}~\cite{wei2019mmgcn} is the first method that constructs a user-item bipartite graph for each modality to obtain modal-specific representations for a better understanding of user preferences.
    \item \textbf{DualGNN}\footnote{https://github.com/wqf321/dualgnn}~\cite{wang2021dualgnn} integrates an additional user co-occurrence graph with the user-item bipartite graph to collaboratively extract the specific fusion pattern for each user.
    \item \textbf{GRCN}\footnote{https://github.com/weiyinwei/GRCN}~\cite{wei2020graph} introduces a graph refinement layer designed to identify noisy edges and eliminate false positives, thus clarifying the structure of the user-item interaction graph.
    \item \textbf{LightGT}\footnote{https://github.com/Liuwq-bit/LightGT}~\cite{wei2023lightgt} employs a transformer-based multimodal recommendation model, develops modal-specific embeddings and a layer-wise position encoder for effective similarity measurement. It also introduces a lightweight self-attention block to enhance the efficiency of self-attention scoring.
    \item \textbf{SLMRec}\footnote{https://github.com/zltao/SLMRec/} \cite{tao2022self} adopts a self-supervised approach using contrastive loss to capture hidden signals within the data, marking a significant evolution in handling multimedia content for personalized recommendations.
    \item \textbf{MMSSL}\footnote{https://github.com/HKUDS/MMSSL}~\cite{wei2023multi} proposes an adversarial self-supervised learning approach with modality-guided collaborative relation generator and discriminator to learn the relations between collaborative signals and multimodal content, with cross-modal contrastive learning to capture dependencies among modality-specific preferences.
\end{itemize}

\subsection{Evaluation Methodology}

\subsubsection{Baselines}
Motivated by \cite{ferrari2019we}, we selected the following traditional recommendation models as baselines. Despite their simplicity, these models have demonstrated superior performance compared to many sophisticated deep learning-based recommendation models~\cite{dong2023newer}. Notably, they make recommendations solely based on user-item interactions.
\begin{itemize}
    \item \textbf{Random}, which randomly selects items for recommendation, without utilizing user preferences or item features. 
    \item \textbf{TopPopular}, which recommends the items that are most popular in the training set.
    \item \textbf{SLIM\_BPR}\footnote{\label{traditional}https://github.com/MaurizioFD/RecSys2019\_DeepLearning\_Evaluation}~\cite{ning2011slim}, which leverages a sparse coefficient matrix to predict user behavior and optimizes its predictions using Bayesian Personalized Ranking (BPR) loss.
    \item \textbf{UserKNN}\footref{traditional}~\cite{ferrari2019we,sarwar2001item}, which is a traditional collaborative filtering method based on the k-nearest neighbors algorithm. We utilize a validation set to select the optimal distance function for user-user similarity from five options: Cosine, Jaccard, Dice, Tversky, and asymmetric distances.
    \item \textbf{ItemKNN}\footref{traditional}~\cite{wang2006unifying}, which is a neighborhood-based method that employs collaborative item-item similarity. We utilize the same approach as UserKNN for tuning the selection of distance functions.
\end{itemize}

\begin{table}[]
\caption{Statistics of the datasets. `\# Inter' refers to the number of user-item interaction. `T Len' and `V Size' refer to the average sentence length and image size, respectively. `-' indicates that the datasets do not provide raw data or the image size is not fixed.}
\resizebox{0.6\columnwidth}{!}{
\centering
\begin{tabular}{ l rrr c c r } 
\toprule
Dataset & \# User & \# Item & \# Inter & T Len & V Size  & Kurtosis  \\
\midrule
\textit{Baby} & 19,445 & 7,050 & 160,792 & 21.74&- & 71.64\\
\textit{Sports} & 35,598 & 18,357 & 296,337  &26.45&- &240.08\\
\textit{Clothing} & 39,387 & 23,033 & 278,677  &20.33&-&94.98\\
\textit{Art} & 25,165 & 9,324 & 201,427 &-&-& 1605.51\\
\textit{Beauty} & 15,576 & 8,678 & 139,318 &-&-& 40.81\\
\textit{Taobao} &  12,539& 8,735 & 83,648 &-&320*320& 150.30 \\
\textit{DY} &  17,876& 5,200 & 124,418 &19.78&300*400&5.99 \\
\bottomrule
\end{tabular}
\label{tab: dataset}
}
\end{table}

\subsubsection{Dataset and Data Preprocessing}
In this study, we follow the original papers in selecting widely recognized and publicly available datasets for e-commerce recommendation, ensuring the reliability of our experimental analysis.
These include the Amazon dataset\footnote{https://jmcauley.ucsd.edu/data/amazon/}~\cite{mcauley2015image}, covering categories such as \textit{Baby}, \textit{Sports\_and\_Outdoors} (short for \textit{Sports}), \textit{Clothing\_Shoes\_and\_Jewelry} (short for \textit{Clothing}), \textit{Arts\_crafts\_and\_Sewing} (short for \textit{Art}), and \textit{Beauty}, as well as datasets from \textit{Taobao}\footnote{https://tianchi.aliyun.com/competition/entrance/231506/information}. Additionally, to evaluate the effectiveness of multimodal information across different recommendation tasks, we utilize the \textit{DY} Dataset~\cite{zhang2024ninerec}, which is specifically tailored for short video recommendations. This diverse selection of datasets enables us to conduct a comprehensive and varied analysis of multimodal recommendation systems.

The statistics of these datasets are summarized in Table~\ref{tab: dataset}. 
`T Len' represents the average length of these text descriptions. `V size' denotes the size of the image, and `-' indicates that the image sizes are not uniform. 
\textit{Clothing} is the largest dataset with the highest number of users and items. And \textit{Sports} contains the longest average text descriptions. 
We assess the Kurtosis for each dataset to evaluate the tail heaviness of the interaction distribution. The kurtosis of a normal distribution is 3, and values greater than this indicate a distribution with a longer tail, suggesting more extreme outliers. For the \textit{DY} dataset, the kurtosis is lower than for the others, indicating a distribution with fewer extreme outliers. This suggests that the dataset is relatively more uniformly distributed.

As for extracting the multimodal features, we preprocess the raw data to obtain latent representations. For the \textit{Baby}, \textit{Sports}, and \textit{Clothing} datasets, titles and text descriptions could be used as textual features. We extract 384-dimensional textual features using pre-trained sentence-transformers~\cite{reimers2019sentence} and adopt the 4096-dimensional visual features following~\cite{zhang2021mining}. User ratings are treated as positive interactions to construct the user-item graph. For \textit{DY}, we follow a similar pipeline, except that image features are extracted using ResNet~\cite{he2016deep}. For \textit{Art}, \textit{Beauty}, and \textit{Taobao}, we directly use the provided interactions and modality features\footref{MGCE}.

\subsubsection{Settings and Metrics}
Different evaluation setups can lead to varying performance results of multimodal recommender, influenced by factors such as data split methods (e.g., leave-one-out, temporal user/global split, random split) and different evaluation metrics~\cite{meng2020exploring,ferrari2019we,malitesta2023disentangling}. To comprehensively analyze model performances and derive reliable conclusions, we first propose reproducing the methods using the same datasets and evaluation procedures as described in the original papers. Subsequently, we will compare all selected methods that are \textit{Code Reproducible} within a uniform setting to ensure consistent and fair benchmarking. 

Among all the selected research works that are both \textit{Code Reproducible} and \textit{Dataset Available}, the majority of studies adopt the following two settings. First, some studies use recall and NDCG as metrics for Top-$N$ recommendations (where $N$ = 10 and 20), employing a random split method to divide the dataset into training, validation, and testing sets in an 8:1:1 ratio. These studies often utilize the Amazon dataset for evaluation, specifically focusing on the three categories: \textit{Baby}, \textit{Sports}, and \textit{Clothing}. Second, other studies opt for a leave-one-out split and randomly select 99 items as negative samples. Performance metrics in these cases include Hit Rate and NDCG for Top-$N$ recommendations (where $N$ = 5, 10). Evaluations often use both \textit{Taobao} dataset and the Amazon dataset (Categories: \textit{Art} and \textit{Beauty}).

As for the uniform setup, we evaluate the recall and NDCG performances of models at $N$ = 10 and 20, using the Amazon \textit{Baby} dataset, the \textit{Taobao} dataset, and the \textit{DY} dataset. We randomly split each dataset into train, validation, and test sets with an 8:1:1 ratio.

\begin{table*}[ht]
\centering
\caption{Experimental results comparing traditional recommendation baselines with multimodal recommendation systems. `Rec@$N$' and `NDCG@$N$' refer to the recall and NDCG performances for the Top-$N$ recommendation. The best results of traditional recommendation baselines are in boldface and gray. Results of multimodal recommendation systems that surpass the best traditional baseline are indicated in green, with darker shades representing better performance. Results that fall below the baseline are shown in red, with darker shades indicating worse performance.}

\label{amazon}
\resizebox{\textwidth}{!}{
\begin{tabular}{lcccc|cccc|cccc}
\toprule
\multicolumn{1}{l}{} & \multicolumn{4}{c}{\textit{Baby}}                                                                                                                   & \multicolumn{4}{c}{\textit{Sports}}                                                                                                               & \multicolumn{4}{c}{\textit{Clothing}}                                                                                       \\ \cmidrule{2-13}
\multirow{-2}{*}{Dataset} & Rec@10                             & NDCG@10                             & Rec@20                             & NDCG@20                             & Rec@10                             & NDCG@10                             & Rec@20                             & NDCG@20                             &Rec@10                             & NDCG@10                             & Rec@20                             & NDCG@20                             \\ \midrule
Random            & 0.0014                           & 0.0006                           & 0.0028                           & 0.0010                            & 0.0007                           & 0.0003                           & 0.0013                           & 0.0005                           & 0.0003                           & 0.0002                           & 0.0010                            & 0.0003                           \\
TopPopular           & 0.0295                           & 0.0161                           & 0.0495                           & 0.0212                           & 0.0180                            & 0.0103                           & 0.0287                           & 0.0131                           & 0.0087                           & 0.0045                           & 0.0150                            & 0.0061                           \\
SLIM BPR             & 0.0545                           & 0.0321                           & 0.0798                           & 0.0387                           & 0.0644                           & 0.0382                           & 0.0875                           & 0.0442                           & 0.0381                           & 0.0227                           & 0.0502                           & 0.0258                           \\
UserKNN              & \cellcolor{gray!50}\textbf{0.0576} & \cellcolor{gray!50}\textbf{0.0328} & \cellcolor{gray!50}\textbf{0.0841} & \cellcolor{gray!50}\textbf{0.0396} & 0.0690                            & 0.0400                             & \cellcolor{gray!50}\textbf{0.0958} & 0.0469                           & 0.0442                           & \cellcolor{gray!50}\textbf{0.0263} & \cellcolor{gray!50}\textbf{0.0599} & \cellcolor{gray!50}\textbf{0.0304} \\
ItemKNN              & 0.0566                           & 0.0327                           & 0.0830                            & 0.0396                           & \cellcolor{gray!50}\textbf{0.0690} & \cellcolor{gray!50}\textbf{0.0415} & 0.0928                           & \cellcolor{gray!50}\textbf{0.0477} & \cellcolor{gray!50}\textbf{0.0443} & 0.0262                           & 0.0593                           & 0.0301                           \\ \midrule
LATTICE              & \cellcolor{red!50}0.0547                           & \cellcolor{red!50}0.0292                           & \cellcolor{green!10}0.0850  & \cellcolor{red!50}0.0370                            & \cellcolor{red!50}0.0620                            & \cellcolor{red!50}0.0335                           & \cellcolor{red!20}0.0953                           & \cellcolor{red!50}0.0421                           & \cellcolor{green!20}0.0492 & \cellcolor{green!10}0.0268 & \cellcolor{green!20}0.0733 & \cellcolor{green!10}0.0330 \\
MICRO                & \cellcolor{red!10}0.0569                           & \cellcolor{red!20}0.0315                           & \cellcolor{green!20}0.0904 & \cellcolor{green!10}0.0401 & \cellcolor{red!10}0.0689                           & \cellcolor{red!40}0.0369                           & \cellcolor{green!20}0.1037 & \cellcolor{red!20}0.0459                           & \cellcolor{green!20}0.0481 & \cellcolor{red!10}0.0256                           & \cellcolor{green!20}0.0704 & \cellcolor{green!10}0.0313 \\
BM3                  & \cellcolor{red!20}0.0564                           & \cellcolor{red!30}0.0301                           & \cellcolor{green!20}0.0883 & \cellcolor{red!20}0.0383                           & \cellcolor{red!30}0.0656                           & \cellcolor{red!40}0.0355                           & \cellcolor{green!10}0.0980 &\cellcolor{red!30}0.0438                           & \cellcolor{red!50}0.0422                           & \cellcolor{red!50}0.0231                           & \cellcolor{green!10}0.0621 & \cellcolor{red!50}0.0281                           \\
FREEDOM              & \cellcolor{green!20}0.0627 & \cellcolor{green!10}0.0330  & \cellcolor{green!40}0.0992 & \cellcolor{green!20}0.0424 & \cellcolor{green!20}0.0717 & \cellcolor{red!30}0.0385                           & \cellcolor{green!40}0.1089 & \cellcolor{green!10}0.0481 & \cellcolor{green!40}0.0629 & \cellcolor{green!40}0.0341 & \cellcolor{green!40}0.0941 & \cellcolor{green!30}0.0420 \\
MGCN                 & \cellcolor{green!10}0.0610  & \cellcolor{red!10}0.0328                           & \cellcolor{green!30}0.0951 & \cellcolor{green!20}0.0416 & \cellcolor{green!30}0.0736 & \cellcolor{red!20}0.0403                           & \cellcolor{green!30}0.1106 & \cellcolor{green!20}0.0499 & \cellcolor{green!50}0.0656 & \cellcolor{green!50}0.0358 & \cellcolor{green!40}0.0961 & \cellcolor{green!40}0.0436 \\
LGMRec               & \cellcolor{green!30}0.0654 & \cellcolor{green!30}0.0353 & \cellcolor{green!40}0.0985 & \cellcolor{green!30}0.0439 & \cellcolor{red!10}0.0690 & \cellcolor{red!30}0.0376                           & \cellcolor{green!20}0.1054 & \cellcolor{red!10}0.0470 & \cellcolor{green!30}0.0554 & \cellcolor{green!30}0.0300 & \cellcolor{green!30}0.0827 & \cellcolor{green!20}0.0369 \\ 
GUME               & \cellcolor{green!50}0.0684 & \cellcolor{green!50}0.0369 & \cellcolor{green!50}0.1040 & \cellcolor{green!50}0.0460 & \cellcolor{green!10}0.0760 & \cellcolor{green!20}0.0419                           & \cellcolor{green!50}0.1148 & \cellcolor{green!50}0.0519 & \cellcolor{green!50}0.0661 & \cellcolor{green!50}0.0359 & \cellcolor{green!50}0.0978 & \cellcolor{green!50}0.0440 \\ 
DiffMM               & \cellcolor{green!10}0.0612 & \cellcolor{red!10}0.0327 & \cellcolor{green!30}0.0933 & \cellcolor{green!10}0.0404 & \cellcolor{green!10}0.0704 & \cellcolor{red!30}0.0380                           & \cellcolor{green!20}0.1066 & \cellcolor{red!10}0.0474 & \cellcolor{green!10}0.0461 & \cellcolor{red!30}0.0244 & \cellcolor{green!20}0.0684 & \cellcolor{red!20}0.0299 \\ 
MENTOR               & \cellcolor{green!30}0.0651 & \cellcolor{green!30}0.0350 & \cellcolor{green!50}0.1027 & \cellcolor{green!40}0.0447 & \cellcolor{green!40}0.0757 & \cellcolor{red!10}0.0411                           & \cellcolor{green!40}0.1129 & \cellcolor{green!30}0.0507 & \cellcolor{green!50}0.0658 & \cellcolor{green!50}0.0357 & \cellcolor{green!50}0.0972 & \cellcolor{green!40}0.0437 \\

\bottomrule
\end{tabular}
}
\end{table*}

\section{Comparative Efficiency}
To assess the efficiency of multimodal recommendation systems, we evaluate them along three key dimensions. First, we compare their performance with traditional models that rely solely on user-item interactions. Second, we assess the impact of removing individual modalities to understand the contribution of each. Third, we examine how performance varies across different levels of interaction sparsity on both the user and item sides, to evaluate the robustness of multimodal models in sparse scenarios.


\subsection{Interaction Only \textit{vs.} Multimodality}
In this section, we compare multimodal recommendation systems with traditional recommendation baselines, using the settings outlined in the original research papers. Furthermore, we utilize a uniform setup to compare all \textit{Code Reproducible} multimodal recommendation models, ensuring consistent comparison and analysis across all methods evaluated.
\subsubsection{Reproducing Original Papers}
\label{sec:reprod}
Table~\ref{amazon} presents experimental results from dividing the dataset into training, validation, and testing sets in an 8:1:1 ratio, while Table~\ref{taobao} displays model performances using a leave-one-out split with 99 randomly selected negative samples.


We make the following observations. Firstly, the top-performing multimodal model usually outperforms the best traditional model in both settings. Notably, as shown in Table~\ref{taobao}, all multimodal recommendation systems surpass the best traditional baseline, highlighting the significant potential of multimodal information to enhance recommendation performance. Secondly, all the reproduced papers that were accepted by the top conferences and journals in the last 5 years apply more up-to-date techniques. These multimodal methods also integrate additional multimodal data, purportedly providing more comprehensive information as outlined in their papers. Counterintuitively, as shown in Table~\ref{amazon}, the simplest KNN methods perform comparably to or even better than many sophisticated multimodal recommendation systems. For example, LATTICE achieves top performance only on the Clothing dataset, while at least two traditional baseline methods outperform it on the Baby and Sports datasets. Similarly, traditional baselines surpass MICRO on at least one evaluation metric. BM3 outperforms traditional models in terms of Recall@20 but underperforms on other metrics. Even the most recent models fail to consistently outperform traditional baselines across all metrics. Notably, in terms of NDCG@10 on the \textit{Sports} dataset, the simplest KNN-based methods rank second among all multimodal approaches. These findings suggest that the current integration of multimodal information in recommendation systems is still suboptimal, underscoring the need for more effective fusion techniques and further research to fully harness the potential of multimodal data.

\begin{table*}[ht]
\centering
\caption{Experimental results comparing traditional recommendation baselines with multimodal recommendation systems. `HR@$N$' and `NDCG@$N$' refer to the hit ratio and NDCG performances for the Top-$N$ recommendation. The best results of traditional recommendation baselines are in boldface and gray. Results of multimodal recommendation systems that surpass the best traditional baseline are indicated in green, with darker shades representing better performance.}
\label{taobao}
\resizebox{\textwidth}{!}{
\begin{tabular}{lcccc|cccc|cccc}
\toprule
\multirow{2}{*}{Dataset} & \multicolumn{4}{c}{\textit{Art}}                                                                                                                   & \multicolumn{4}{c}{\textit{Beauty}}                                                                                                                & \multicolumn{4}{c}{\textit{Taobao}}                                                                                                                \\ \cmidrule{2-13}
                         & HR@5                             & NDCG@5                              & HR@10                            & NDCG@10                             & HR@5                             & NDCG@5                              & HR@10                            & NDCG@10                             & HR@5                             & NDCG@5                              & HR@10                            & NDCG@10                             \\ \midrule
Random                & 0.0408                           & 0.0239                           & 0.0788                           & 0.0361                           & 0.0494                           & 0.0296                           & 0.0991                           & 0.0456                           & 0.0525                           & 0.0307                           & 0.1028                           & 0.0468                           \\
TopPopular               & 0.2165                           & 0.1512                           & 0.3067                           & 0.1802                           & 0.2313                           & 0.1581                           & 0.3530                            & 0.1971                           & 0.2374                           & 0.1601                           & 0.3456                           & 0.1952                           \\
SLIM BPR                 & 0.3913                           & 0.3260                            & 0.4567                           & 0.3472                           & 0.4381                           & 0.3483                           & \cellcolor{gray!50}\textbf{0.5296} & 0.3780                            & 0.3068                           & 0.2359                           & 0.3795                           & 0.2594                           \\
UserKNN                  & \cellcolor{gray!50}\textbf{0.3967} & 0.3290                            & 0.4553                           & 0.3480                            & \cellcolor{gray!50}\textbf{0.4441} & \cellcolor{gray!50}\textbf{0.3545} & 0.5243                           & \cellcolor{gray!50}\textbf{0.3804} & 0.3105                           & 0.2458                           & 0.3732                           & 0.2661                           \\
ItemKNN                  & 0.3942                           & \cellcolor{gray!50}\textbf{0.3302} & \cellcolor{gray!50}\textbf{0.4568} & \cellcolor{gray!50}\textbf{0.3504} & 0.4149                           & 0.3318                           & 0.4816	                           & 0.3535                           & \cellcolor{gray!50}\textbf{0.3157} & \cellcolor{gray!50}\textbf{0.2489} & \cellcolor{gray!50}\textbf{0.3912} & \cellcolor{gray!50}\textbf{0.2731} \\ \midrule
MGCL                     & \cellcolor{green!20}0.7112 & \cellcolor{green!40}0.6107 & \cellcolor{green!20}0.7928 & \cellcolor{green!30}0.6371 & \cellcolor{green!20}0.5571 & \cellcolor{green!20}0.4333 & \cellcolor{green!30}0.6636 & \cellcolor{green!20}0.4680 & \cellcolor{green!20}0.3863 & \cellcolor{green!20}0.2907 & \cellcolor{green!20}0.4920 & \cellcolor{green!20}0.3246 \\
MCLN                     & \cellcolor{green!30}0.7146 & \cellcolor{green!20}0.6071 & \cellcolor{green!40}0.7977 & \cellcolor{green!20}0.6340  & \cellcolor{green!40}0.5623 & \cellcolor{green!30}0.4341 & \cellcolor{green!50}0.6688 & \cellcolor{green!30}0.4686 & \cellcolor{green!40}0.3973 & \cellcolor{green!40}0.3021 & \cellcolor{green!30}0.5002 & \cellcolor{green!40}0.3353 \\
MGCE                     & \cellcolor{green!50}0.7211 & \cellcolor{green!50}0.6213 & \cellcolor{green!50}0.7986 & \cellcolor{green!50}0.6465 & \cellcolor{green!50}0.5701 & \cellcolor{green!50}0.4473 & \cellcolor{green!40}0.6683 & \cellcolor{green!50}0.4792 & \cellcolor{green!50}0.4120 & \cellcolor{green!50}0.3082 & \cellcolor{green!50}0.5231 & \cellcolor{green!50}0.3440 \\ \bottomrule
\end{tabular}
}
\end{table*}

\begin{table*}
\centering
\caption{Experimental results comparing traditional recommendation baselines, \textit{Reproducible} multimodal recommendation systems, and multimodal recommendation systems that are \textit{Code Reproducible} only. `Rec@$N$' and `NDCG@$N$' refer to the recall and NDCG performances for the Top-$N$ recommendation. `-' indicates that the model did not converge. The best results from traditional recommendation baselines are in boldface and gray. Results of multimodal recommendation systems that surpass the best traditional baseline are indicated in green, with darker shades representing better performance. Results that fall below the baseline are shown in red, with darker shades indicating worse performance.}
\resizebox{\textwidth}{!}{
\begin{tabular}{lcccc|cccc|cccc}
\toprule
\multirow{2}{*}{Model} & \multicolumn{4}{c}{\textit{Baby}} & \multicolumn{4}{c}{\textit{Taobao}} & \multicolumn{4}{c}{\textit{DY}} \\ \cmidrule{2-13}
 & Rec@10 & NDCG@10 & Rec@20 & NDCG@20 & Rec@10 & NDCG@10 & Rec@20 & NDCG@20 & Rec@10 & NDCG@10 & Rec@20 & NDCG@20 \\ \midrule
ItemKNN&0.0566&0.0327&0.0830&0.0396&0.0554&0.0263&\cellcolor{gray!50}\textbf{0.0920}&0.0354&0.2920&0.1960&0.3477&0.2102\\
UserKNN&\cellcolor{gray!50}\textbf{0.0576}&\cellcolor{gray!50}\textbf{0.0328}&\cellcolor{gray!50}\textbf{0.0841}&\cellcolor{gray!50}\textbf{0.0396}&\cellcolor{gray!50}\textbf{0.0580}&\cellcolor{gray!50}\textbf{0.0277}&0.0908&\cellcolor{gray!50}\textbf{0.0360}&\cellcolor{gray!50}\textbf{0.2953}&\cellcolor{gray!50}\textbf{0.2000}&\cellcolor{gray!50}\textbf{0.3488}&\cellcolor{gray!50}\textbf{0.2138}\\\midrule
LATTICE   & \cellcolor{red!20}0.0547 & \cellcolor{red!20}0.0292 & \cellcolor{green!10}0.0850 & \cellcolor{red!10}0.0370 & - & - & - & - & \cellcolor{red!30}0.2491 & \cellcolor{red!30}0.1533 & \cellcolor{red!20}0.3247 & \cellcolor{red!20}0.1726 \\
MICRO     & \cellcolor{red!10}0.0569 & \cellcolor{red!10}0.0315 & \cellcolor{green!20}0.0904 & \cellcolor{green!10}0.0401 & - & - & - & - &\cellcolor{red!40} 0.2231 & \cellcolor{red!40}0.1332 & \cellcolor{red!40}0.2955 & \cellcolor{red!40}0.1517 \\
BM3       & \cellcolor{red!10}0.0564 & \cellcolor{red!20}0.0301 & \cellcolor{green!20}0.0883 & \cellcolor{red!10}0.0383 & \cellcolor{red!40}0.0461 & \cellcolor{red!50}0.0189 & \cellcolor{red!40}0.0786 & \cellcolor{red!40}0.0270 & \cellcolor{red!40}0.2026 & \cellcolor{red!40}0.1199 & \cellcolor{red!40}0.2831 & \cellcolor{red!40}0.1405 \\
FREEDOM   & \cellcolor{green!20}0.0627 & \cellcolor{green!10}0.0330 & \cellcolor{green!30}0.0992 & \cellcolor{green!20}0.0424 & \cellcolor{red!40}0.0439 & \cellcolor{red!50}0.0187 & \cellcolor{red!40}0.0776 & \cellcolor{red!40}0.0271 & \cellcolor{red!40}0.2162 & \cellcolor{red!40}0.1299 & \cellcolor{red!40}0.2874 & \cellcolor{red!40}0.1481 \\
MGCN      & \cellcolor{green!10}0.0610 & \cellcolor{red!10}0.0328 & \cellcolor{green!30}0.0951 & \cellcolor{green!20}0.0416 & - & - & - & - & \cellcolor{red!30}0.2499 & \cellcolor{red!30}0.1523 & \cellcolor{red!20}0.3221 & \cellcolor{red!20}0.1708 \\
LGMRec    & \cellcolor{green!30}0.0654 & \cellcolor{green!20}0.0353 & \cellcolor{green!30}0.0985 & \cellcolor{green!30}0.0439 & \cellcolor{red!30}0.0490 & \cellcolor{red!30}0.0217 & \cellcolor{red!20}0.0857 & \cellcolor{red!30}0.0309 & \cellcolor{red!30}0.2439 & \cellcolor{red!30}0.1506 & \cellcolor{red!20}0.3144 & \cellcolor{red!20}0.1686 \\
MGCL      & \cellcolor{green!40}0.0678 & \cellcolor{green!40}0.0401 & \cellcolor{green!40}0.1027 & \cellcolor{green!40}0.0499 & \cellcolor{green!10}0.0583 & \cellcolor{red!10}0.0275 & \cellcolor{green!20}0.0974 & \cellcolor{green!30}0.0373 & \cellcolor{red!10}0.2924 & \cellcolor{red!10}0.1961 & \cellcolor{green!30}0.3667 & \cellcolor{green!20}0.2159 \\
MCLN      & \cellcolor{green!40}0.0684 & \cellcolor{green!30}0.0392 & \cellcolor{green!40}0.1028 & \cellcolor{green!40}0.0487 & \cellcolor{red!10}0.0574 & \cellcolor{red!10}0.0254 & \cellcolor{green!50}0.1039 & \cellcolor{green!10}0.0369 & \cellcolor{red!30}0.2306 & \cellcolor{red!30}0.1505 & \cellcolor{red!30}0.3074 & \cellcolor{red!20}0.1709 \\
MGCE      & \cellcolor{green!50}0.0720 & \cellcolor{green!50}0.0421 & \cellcolor{green!50}0.1100 & \cellcolor{green!50}0.0527 & \cellcolor{green!50}0.0612 & \cellcolor{green!10}0.0278 & \cellcolor{green!40}0.1027 & \cellcolor{green!50}0.0381 & \cellcolor{green!50}0.3062 & \cellcolor{green!50}0.2074 & \cellcolor{green!50}0.3777 & \cellcolor{green!50}0.2267 \\ 
GUME               & \cellcolor{green!40}0.0684 & \cellcolor{green!20}0.0369 & \cellcolor{green!40}0.1040 & \cellcolor{green!30}0.0460 & - & - & - & - & \cellcolor{red!10}0.2711 & \cellcolor{red!10}0.1712 & \cellcolor{red!10}0.3383 & \cellcolor{red!10}0.1884 \\ 
DiffMM               & \cellcolor{green!10}0.0612 & \cellcolor{red!10}0.0327 & \cellcolor{green!30}0.0933 & \cellcolor{green!10}0.0404 & \cellcolor{red!40}0.0490 & \cellcolor{red!20}0.0220                           & \cellcolor{red!20}0.0872 & \cellcolor{red!20}0.0314 & \cellcolor{red!40}0.2244 & \cellcolor{red!40}0.1359 & \cellcolor{red!30}0.2982 & \cellcolor{red!40}0.1548 \\ 
MENTOR               & \cellcolor{green!30}0.0651 & \cellcolor{green!20}0.0350 & \cellcolor{green!40}0.1027 & \cellcolor{green!30}0.0447 & \cellcolor{red!30}0.0502 & \cellcolor{red!20}0.0226                           & \cellcolor{red!10}0.0891 & \cellcolor{red!10}0.0322 & \cellcolor{red!30}0.2416 & \cellcolor{red!30}0.1496 & \cellcolor{red!30}0.3068 & \cellcolor{red!30}0.1663 \\ 

\midrule
VBPR      & \cellcolor{red!40}0.0423 & \cellcolor{red!40}0.0223 & \cellcolor{red!40}0.0663 & \cellcolor{red!40}0.0284 & \cellcolor{red!30}0.0494 & \cellcolor{red!20}0.0237 & \cellcolor{red!30}0.0817 & \cellcolor{red!20}0.0318 & \cellcolor{red!30}0.2478 & \cellcolor{red!30}0.1519 &\cellcolor{red!20} 0.3211 & \cellcolor{red!20}0.1706 \\
MMGCN     & \cellcolor{red!50}0.0378 & \cellcolor{red!50}0.0200 & \cellcolor{red!50}0.0615 & \cellcolor{red!50}0.0261 & \cellcolor{red!50}0.0396 & \cellcolor{red!50}0.0184 & \cellcolor{red!50}0.0698 & \cellcolor{red!50}0.0259 & \cellcolor{red!50}0.1269 & \cellcolor{red!50}0.0696 & \cellcolor{red!50}0.1882 & \cellcolor{red!50}0.0853 \\
GRCN      & \cellcolor{red!20}0.0539 & \cellcolor{red!20}0.0288 & \cellcolor{red!10}0.0833 & \cellcolor{red!20}0.0363 & \cellcolor{red!20}0.0550 & \cellcolor{green!30}0.0283 & \cellcolor{red!10}0.0890 & \cellcolor{green!10}0.0368 & \cellcolor{red!20}0.2650 & \cellcolor{red!20}0.1671 &\cellcolor{red!10}0.3359 & \cellcolor{red!10}0.1853 \\
DualGNN   & \cellcolor{red!40}0.0448 & \cellcolor{red!30}0.0240 & \cellcolor{red!30}0.0716 & \cellcolor{red!30}0.0309 & \cellcolor{red!10}0.0570 & \cellcolor{green!10}0.0279 & \cellcolor{green!20}0.0973 & \cellcolor{green!50}0.0380 & \cellcolor{red!30}0.2384 & \cellcolor{red!30}0.1474 & \cellcolor{red!30}0.3088 & \cellcolor{red!30}0.1654 \\
SLMRec    & \cellcolor{red!20}0.0529 & \cellcolor{red!20}0.0290 & \cellcolor{red!20}0.0775 & \cellcolor{red!20}0.0353 & \cellcolor{red!20}0.0518 & \cellcolor{red!20}0.0230 & \cellcolor{red!20}0.0858 & \cellcolor{red!20}0.0315 & \cellcolor{red!20}0.2568 & \cellcolor{red!20}0.1580 & \cellcolor{red!10}0.3316 & \cellcolor{red!10}0.1771 \\

LightGT   & \cellcolor{red!30}0.0477 &\cellcolor{red!30}0.0250 & \cellcolor{red!20}0.0753 & \cellcolor{red!30}0.0314 & \cellcolor{red!50}0.0411 & \cellcolor{red!50}0.0186 & \cellcolor{red!20}0.0845 & \cellcolor{red!30}0.0304 & \cellcolor{red!50}0.1119 & \cellcolor{red!50}0.0595 & \cellcolor{red!50}0.1693 & \cellcolor{red!50}0.0745 \\
MMSSL               & \cellcolor{green!20}0.0629 & \cellcolor{green!20}0.0353 & \cellcolor{green!30}0.0948 & \cellcolor{green!30}0.0441 & \cellcolor{red!40}0.0485 & \cellcolor{red!30}0.0210                           & \cellcolor{red!10}0.0898 & \cellcolor{red!20}0.0313 & \cellcolor{red!20}0.2525 & \cellcolor{red!20}0.1574 & \cellcolor{red!20}0.3245 & \cellcolor{red!10}0.1760 \\ 
\bottomrule
\end{tabular} 
\label{same setting}
}
\end{table*}

\subsubsection{Uniform Benchmarking}
To ensure a consistent comparison across all baselines, we evaluate all \textit{Code Reproducible} multimodal models under the same settings and benchmark them against traditional models. Our evaluation utilizes various datasets, including Amazon \textit{Baby} dataset, \textit{Taobao} dataset, the \textit{DY} dataset.


As shown in Table~\ref{same setting}, our reproduced experiments support the findings reported in Section~\ref{sec:reprod}. Traditional KNN methods consistently outperform many more complex multimodal models, underscoring the challenges of effectively leveraging multimodal data. On the \textit{DY} dataset, for instance, KNN outperforms most multimodal approaches, likely due to data quality issues or misalignment between visual and textual contents. 
While multimodal models are designed to integrate diverse data sources to enhance recommendation performance, their ability to leverage complementary information is not always guaranteed, particularly when feature representations across modalities are misaligned or when redundant or noisy features obscure meaningful signals~\cite{liu2023multimodal}. In contrast, KNN relies directly on user-item interactions, preserving the core collaborative filtering signal. These results suggest that increasing model complexity does not necessarily lead to better performance, highlighting the need for more robust multimodal fusion strategies.

\begin{table*}
    \centering
    \caption{Experimental results demonstrating the impact of each modality on recommendation performance, measured by Recall@10. The table compares the original multimodal model (`Original'), models with either text (`w/o T') or visual (`w/o V') modalities removed, and model uses only user-item interactions (`Interaction Only'). The best results among all multimodal models are highlighted in gray, while the best performance between the original model and its variants is shown in boldface. `Ablation study' indicates whether the original paper conducted experiments to assess the impact of removing single modalities on recommendation accuracy.}
    \resizebox{0.95\textwidth}{!}{
    \begin{tabular}{lccccc|cc|cccc}
\toprule
\multirow{3}{*}{Model}&\multirow{3}{*}{\begin{tabular}[c]{@{}c@{}}Ablation\\ Study\end{tabular}} & \multicolumn{4}{c}{\textit{Baby}} & \multicolumn{2}{c}{\textit{Taobao}} & \multicolumn{4}{c}{\textit{DY}} \\
\cmidrule{3-12}
 && \textbf{w/o T} & \textbf{w/o V} & \textbf{Original} & \begin{tabular}[c]{@{}c@{}}\textbf{Interaction}\\ \textbf{Only}\end{tabular} & \textbf{w/o V} & \textbf{Original} & \textbf{w/o T} & \textbf{w/o V} & \textbf{Original} & \begin{tabular}[c]{@{}c@{}}\textbf{Interaction}\\ \textbf{Only}\end{tabular} \\ \midrule

        VBPR & \xmark & \textbf{0.0428} & 0.0400 & 0.0423 & 0.0386 & \textbf{0.0495} & 0.0494 & \textbf{0.2547} & 0.2426 & 0.2478 & 0.2510 \\ 
        MMGCN & \cmark & \textbf{0.0384} & 0.0365 & 0.0378 & 0.0342 & \textbf{0.0545} & 0.0396 & 0.1203 & 0.1154 & \textbf{0.1269} & 0.1199 \\ 
        GRCN & \cmark & 0.0488 & 0.0517 & \textbf{0.0539} & 0.0485 & \cellcolor{gray!50}\textbf{0.0567} & 0.0550 & 0.2435 & 0.2367 & 0.2650 & \textbf{0.2692}  \\ 
        DualGNN & \cmark & 0.0511 & \textbf{0.0612} & 0.0448 & 0.0377 & 0.0329 & \textbf{0.0570} & 0.2430 & 0.2402 & 0.2384 & \textbf{0.2534}  \\ 
        LATTICE & \xmark & 0.0492 & 0.0546 & \textbf{0.0547} & 0.0469 & - & - & \textbf{0.2544} & 0.2515 & 0.2491 & 0.2484  \\ 
        MICRO & \xmark & 0.0487 & \textbf{0.0580} & 0.0569 & 0.0409 & - & - & 0.2304 & 0.2348 & 0.2231 & \textbf{0.2393}  \\
        SLMRec* & \xmark & 0.0475 & 0.0495 & \textbf{0.0529} & 0.0476 & \textbf{0.0548} & 0.0518 & 0.2542 & \textbf{0.2594} & 0.2568 & 0.2544  \\ 
        BM3 & \cmark & 0.0544 & \textbf{0.0571} & 0.0564 & 0.0561 & \textbf{0.0476} & 0.0461 & 0.2078 & 0.2006 & 0.2026 & \textbf{0.2082}  \\ 
        FREEDOM & \cmark & 0.0501 & 0.0622 & \textbf{0.0627} & 0.0443 & 0.0412 & \textbf{0.0439} & \textbf{0.2228} & 0.2119 & 0.2162 & 0.2226 \\ 
        MMSSL* & \xmark & 0.0507 & 0.0613 & \textbf{0.0629} & 0.0462 & \textbf{0.0525} & 0.0485 & 0.2505 & 0.2470 & \textbf{0.2525} & 0.2489 \\ 
        LightGT & \xmark & 0.0394 & 0.0421 & \textbf{0.0477} & 0.0331 & 0.0281 & \textbf{0.0411} & \textbf{0.2069} & 0.0964 & 0.1119 & 0.0670  \\ 
        MGCN & \cmark & 0.0528 & \textbf{0.0640} & 0.0610 & 0.0486 & - & - & 0.2249 & 0.2291 & 0.2499 & \textbf{0.2585}  \\ 
        MGCL* & \cmark & 0.0613 & 0.0663 & \textbf{0.0678} & 0.0569 & 0.0441 & \textbf{0.0583} & 0.2817 & 0.2886 & \textbf{0.2924} & 0.1940 \\ 
        MCLN & \cmark & \cellcolor{gray!50}0.0637 & \textbf{0.0699} & 0.0684 & 0.0461 & 0.0443 & \textbf{0.0574} & \textbf{0.2686} & 0.2576 & 0.2306 & 0.1969  \\ 
        MGCE* & \cmark & 0.0634 & \cellcolor{gray!50}0.0711 & \cellcolor{gray!50}\textbf{0.0720} & \cellcolor{gray!50}0.0607 & 0.0537 & \cellcolor{gray!50}\textbf{0.0612} & \cellcolor{gray!50}0.3129 & \cellcolor{gray!50}\textbf{0.3130} & \cellcolor{gray!50}0.3062 & \cellcolor{gray!50}0.2785  \\ 
        LGMRec* & \xmark & 0.0499 & 0.0615 & \textbf{0.0654} & 0.0395 & \textbf{0.0505} & 0.0490 & 0.2405 & \textbf{0.2441} & 0.2439 & 0.2368 \\ 
        GUME* & \xmark & 0.0556 & 0.0597 & \textbf{0.0684} & 0.0523 & - & - & 0.2643 & 0.2730 & 0.2711 & \textbf{0.2741} \\ 
        DiffMM* & \xmark & 0.0533 & 0.0592 & \textbf{0.0612} & 0.0520 & 0.0475 & \textbf{0.0490} & 0.2273 & 0.2337 & 0.2244 & \textbf{0.2381} \\ 
        MENTOR* & \xmark & 0.0510 & \textbf{0.0668} & 0.0651 & 0.0487 & 0.0479 & \textbf{0.0502} & 0.2543 & 0.2450 & 0.2416 & \textbf{0.2596} \\ 
        \bottomrule
    \end{tabular}
    }
    \label{modality missing}
\end{table*}

\subsection{Multimodality \textit{vs.} Single Modality}
Understanding the impact of each modality is essential for effectively utilizing multimodal information and improving model performance. To evaluate this, we assess how recommendation performance changes when a specific modality is removed or replaced with a randomly initialized input. This reveals the contribution of each modality to the overall system.

As shown in Table~\ref{modality missing}\footnote{All models are listed in order of publication year.}, we include all \textit{Code Reproducible} models and construct variants by removing either visual, textual, or all modalities, keeping only user-item interactions. The * in the table means randomly initialized input. We then compare these variants against the original multimodal models. Additionally, we review each paper to check whether similar ablation studies about modality were reported.



Intuitively, incorporating additional multimodal information should improve a recommendation system's ability to model user preferences and enhance accuracy. However, our experimental results fall short of this expectation. First, the original multimodal systems do not consistently outperform their ablated variants. On both the \textit{Baby} and \textit{Taobao} datasets, only about half of the models outperform their variants, and on the \textit{DY} dataset, just 3 out of 19 models show superior performance. Moreover, even for models where multimodal inputs lead to the best results, the performance gains over their strongest unimodal variants are often marginal, calling into question the practical benefit of adding complex modalities. This suggests that simply incorporating more modalities does not guarantee better user modeling or recommendation quality. 

Furthermore, while half of the research works conduct ablation studies on individual modalities in their original papers, the experimental results do not always align with ours. This inconsistency may stem from variations in the dataset and experimental settings. 


\begin{figure}[ht]
    \centering
    \includegraphics[width=0.7\linewidth]{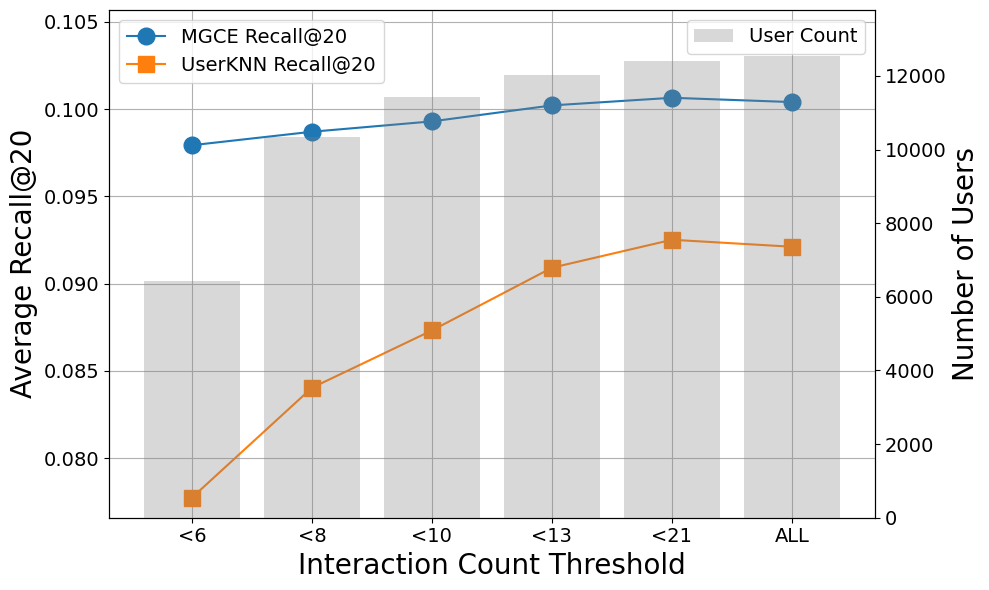}
    \caption{Performance \textit{w.r.t} user interaction frequencies. 
Right side \textit{y}-axis: \# of users included; Left side \textit{y}-axis: average performance measured by Recall@20.}
    \label{fig:user_interaction}
\end{figure}

\subsection{Effect of Interaction Sparsity on Model Performance}

User-item interactions in real world recommender systems are inherently sparse. The items suffer from the long-tail distribution and users also vary significantly in their interaction density. This sparsity poses a fundamental challenge for collaborative filtering based models, which rely solely on interactions to learn user preferences. In particular, users with fewer historical interactions often receive less accurate or unreliable recommendations due to insufficient collaborative signals.

One of the key motivations for multimodal recommendation systems is their potential to mitigate data sparsity by leveraging additional item-side information such as text and image. These modality features are expected to compensate for limited user-item interactions, enabling better modeling of user preferences. However, it remains unclear whether multimodal models consistently outperform traditional methods under sparse conditions. To explore this, we conduct an ablation study to examine how interaction sparsity affects the effectiveness of multimodal recommendations. By comparing performance under varying levels of interaction sparsity on user and item sides, we aim to determine whether the benefit of multimodal information is more pronounced when interaction is sparse.

To investigate this, we compare the performance of a strong multimodal model, MGCE (which achieved top performance in our earlier study), with a traditional collaborative filtering baseline, UserKNN, under varying degrees of interaction sparsity. We evaluate both user-side and item-side sparsity conditions.

Figure~\ref{fig:user_interaction} shows the average Recall@20 of MGCE and UserKNN grouped by user interaction thresholds (e.g., fewer than 6, 8, 10 interactions, etc.). MGCE consistently outperforms UserKNN across all sparsity levels. The performance gap is most pronounced in the lowest interaction group, where UserKNN struggles due to limited collaborative signals. This highlights the value of modality information in sparse settings. As interaction counts increase, both models achieve higher recall, though the relative advantage of MGCE gradually narrows. This trend underscores the importance of interaction data for traditional models while demonstrating that multimodal features are especially beneficial in low-data scenarios. Additionally, the user distribution (gray bars) reveals that a large portion of users fall into sparse interaction groups, emphasizing the practical need to optimize recommendation models using modality data. 

\begin{figure}[ht]
    \centering
    \includegraphics[width=\linewidth]{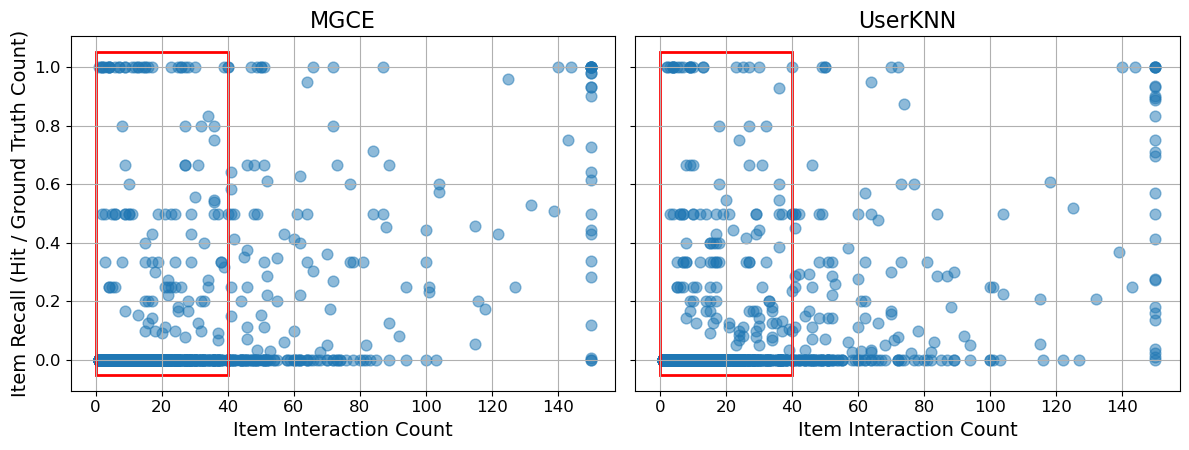}
    \caption{Performance \textit{w.r.t} item sparsity degrees. 
Each point represents item-specific recommendation accuracy among the test set, calculated by the number of correctly recommended items over the number of occurrences in the test set.}
    \label{fig:item_interaction}
\end{figure}

Figure~\ref{fig:item_interaction} illustrates model performance across varying item interaction counts, with each point representing an individual item. Performance is calculated as the ratio of correctly recommended items to the number of times the item appears in the test set. The red box highlights the low-interaction region (fewer than 40 interactions), where UserKNN exhibits noticeably lower performance. In this region, many points cluster near the bottom, reflecting the difficulty KNN faces in generating accurate recommendations without sufficient collaborative data. In contrast, MGCE enables more items to achieve high recall, demonstrating that modality features can help improve recommendation quality for items with sparse interaction histories.

These findings prove the necessity of incorporating multimodal signals into recommendation models, particularly to address data sparsity problem. The resilience of multimodal recommendation models to sparsity makes them more suitable for large-scale and real-world scenarios where interactions are often limited.

\section{Recommendation Tasks}
The impact of different types of modal information on user decision-making can vary across recommendation scenarios. To better understand this, we conduct a study on how multimodal data contributes to model performance across diverse recommendation tasks, focusing specifically on e-commerce and short video recommendations. This analysis helps reveal whether certain modalities are more effective in particular domains, offering insights for task-specific model design. We also include a literature review in Section~\ref{sec:otherrectasks} to explore how the importance of modality varies between other different recommendations tasks, such as news, music, food and POI recommendations, providing a more comprehensive view.

As illustrated in Table~\ref{modality missing}, different types of modal information contribute differently depending on the recommendation scenario. In e-commerce settings, textual features often play a more important role. For example, in the \textit{Baby} dataset, although 11 models achieve their best performance with multimodal inputs, 6 of them show only small improvements over using text alone, indicating limited benefit from adding visual features. In contrast, for short video recommendations, visual information tends to be more useful. In the \textit{DY} dataset, 11 out of 19 models using only image features outperform their original multimodal versions. This difference likely stems from the nature of short-video platforms, where visual content is more closely tied to user preferences. These findings suggest that the effectiveness of each modality depends on the task, highlighting the need for tailored strategies to better leverage multimodal information in different domains.

\section{Recommendation Stages}

In this section, our objective is to explore whether the impact of multimodal information remains the same across different stages of the recommendation process. We focus on two key stages: recall and reranking. Recall is designed to filter and retrieve all potentially relevant items from a vast item pool. Reranking, on the other hand, specifically addresses the order of these retrieved items, positioning the most relevant items at the top of the recommendation list. 

We evaluate the model performance for the Top-$N$ recommendations with varying $N$ values, $N \in \{1,5,10,20,50,100\}$. We hypothesize that with a smaller $N$, the model's performance more closely reflects that of the reranking stage of the recommendation process. Conversely, with a larger $N$, the model's performance aligns more with the recall stage.

\begin{figure*}[ht]
    \centering
    \begin{minipage}[b]{0.33\textwidth}
        \centering
        \includegraphics[width=\textwidth]{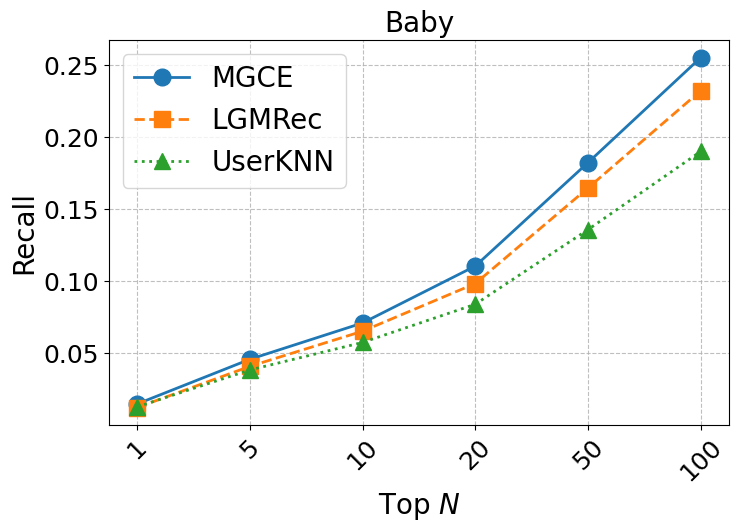}
    \end{minipage}
    \begin{minipage}[b]{0.33\textwidth}
        \centering
        \includegraphics[width=\textwidth]{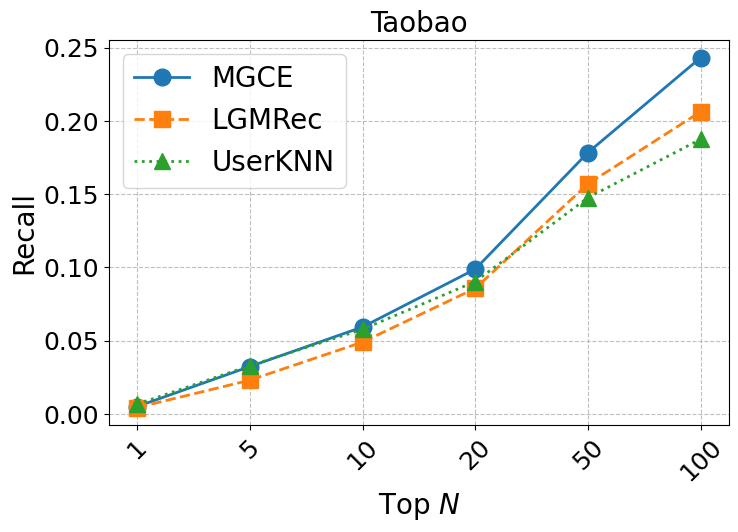}
    \end{minipage}
    \begin{minipage}[b]{0.33\textwidth}
        \centering
        \includegraphics[width=\textwidth]{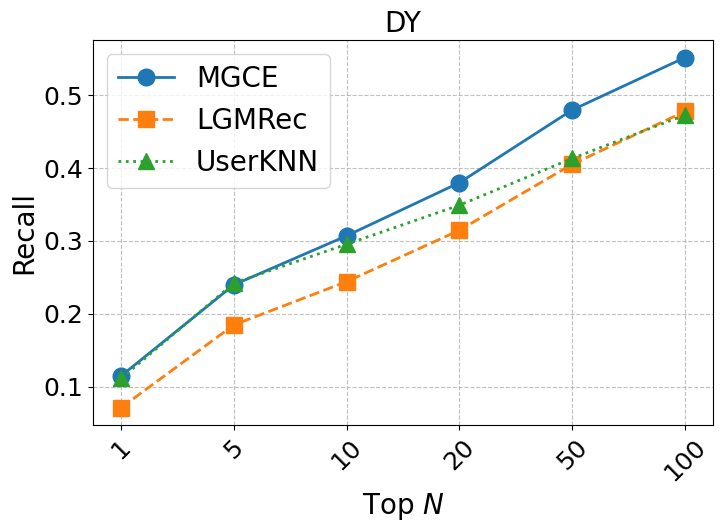}
    \end{minipage}

    \caption{Recall performance of multimodal models MGCE and LGMRec, and the traditional model UserKNN, for Top-$N$ recommendation where $N$ $\in \{1,5,10,20,50,100\}$, on the \textit{Baby}, \textit{Taobao}, and \textit{DY} datasets.}
    \label{Rerank and Recall}
\end{figure*}

As shown in Figure~\ref{Rerank and Recall}, we compare the performance of two leading multimodal recommendation systems, MGCE and LGMRec, against the top traditional method, UserKNN. We observed that the performance improvement of multimodal models increases with larger values of $N$. On the \textit{Baby} dataset, the relative improvement of MGCE over UserKNN increases from 15.10\% to 34.23\% as $N$ increases from 1 to 100 in terms of recall. Similarly, the relative improvement of LGMRec over UserKNN rises from -3.88\% to 22.13\%. 

These results indicate that multimodal information tends to be more beneficial in the recall stage than in the reranking stage. More specifically, multimodal information enhances the model's ability to select promising candidate items from a large pool that users may prefer. However, it may have limited effectiveness in accurately modeling user preferences or could even mislead the model and introduce noise. When $N$=1 or $N$=5, the traditional UserKNN algorithm performs comparably to the multimodal algorithm and even significantly outperforms LGMRec on the \textit{DY} dataset.

\begin{figure}
    \centering
    \includegraphics[width=0.5\linewidth]{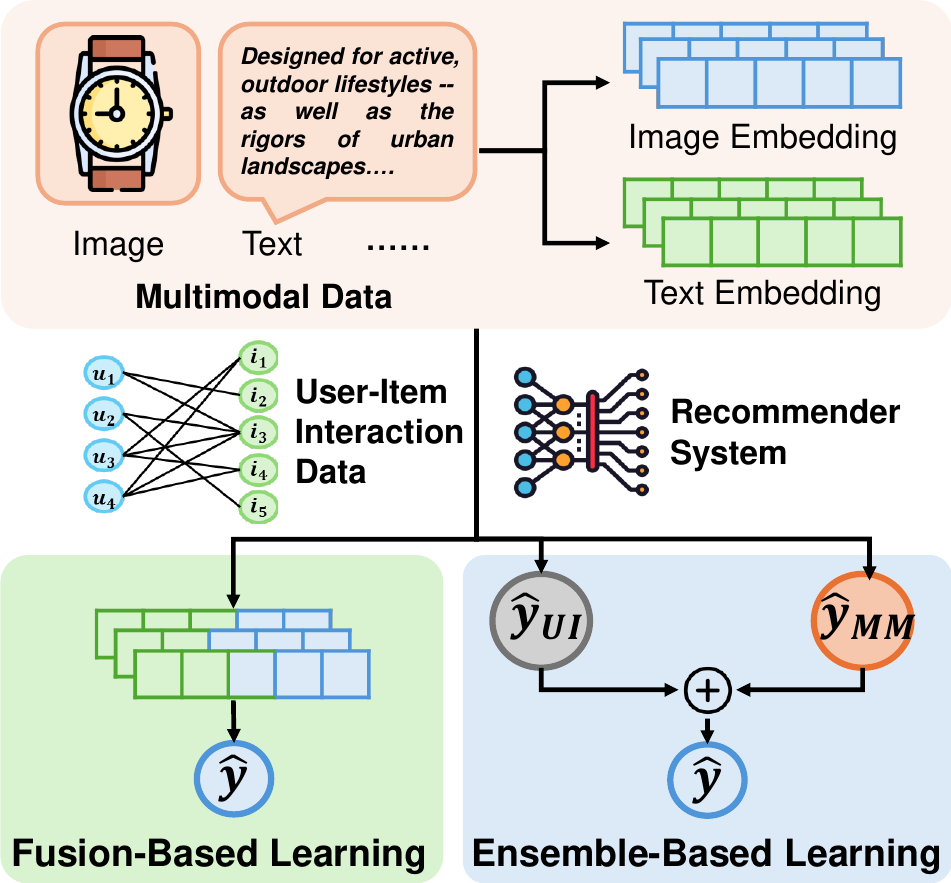}
    \caption{Approaches for learning multimodal features, including Fusion-Based and Ensemble-Based Learning.}
    \label{model structure}
\end{figure}

\section{Multimodal Data Integration}

Model architecture plays a critical role in effectively leveraging multimodal data for recommendation systems~\cite{liu2024multimodal}. In this work, we analyze a range of multimodal recommendation models to understand how different architectural designs impact performance, particularly in how multimodal features are integrated with collaborative filtering signals from user-item interactions.

From the results in Table~\ref{same setting} and Table~\ref{modality missing}, we observe that three models, MGCL, MCLN, and MGCE, not only outperform other multimodal approaches but also show stronger robustness when a single modality is removed. MGCE without modality feature outperforms models such as LATTICE, BM3, and MGCN using modality information. These results suggest that MGCE is more effective at learning collaborative embeddings, and the learned representations can be more informative when collaborative signals integrate with modality data. In contrast, the other inferior models fail to capture strong collaborative signals, and the inclusion of modality data contributes little or even introduces noise. To understand the reasons behind their superior performance, we examine their model architectures in more detail.

We find that most multimodal recommendation models follow a common structure: they first extract multimodal features from raw modality data and then refine the modality embeddings using collaborative signals or the homogeneous relations to predict user preferences. As illustrated in Figure~\ref{model structure}, these models can generally be grouped into two categories based on how they combine multimodal and interaction information: Fusion-Based Learning and Ensemble-Based Learning. Fusion-based methods generate a unified embedding by merging modality features and interaction data early in the pipeline. In contrast, ensemble-based methods produce separate predictions from each source and combine them at the final stage.

Interestingly, all three top-performing models, MGCL, MCLN, and MGCE, adopt the Ensemble-Based Learning approach, while the remaining models rely on Fusion-Based Learning. Several factors may explain the advantage of ensemble-based methods. First, by handling each modality independently, they allow the model to learn distinct preference signals without early interference, reducing the risk of noise amplification from less relevant or low-quality modalities. Second, ensemble-based designs provide greater flexibility in balancing and dynamically adjusting the influence of different supervision signals, which can lead to better adaptability and overall performance.

Among the three top-performing models, MGCE achieves the best overall performance. This may be attributed to its incorporation of item popularity. MGCE introduces a popularity-aware graph Laplacian norm that embeds users’ sensitivity to popularity into the representation learning process. In addition, it treats popularity as one item attribute, integrating it into the multimodal conformity learning framework. This design highlights the critical role of popularity in recommendation systems. Popularity signals often reflect collective user preferences and can enhance recommendation accuracy, especially in cold-start or sparse data scenarios. By modeling user-specific sensitivity to popularity, MGCE balances personalized preferences with global trends, leading to more nuanced recommendations. 

\section{Discussion}

\subsection{Beyond E-Commerce and Short Video: How does Modality Importance Vary across Recommendation Tasks?}
\label{sec:otherrectasks}

To gain a more comprehensive understanding of how different modalities contribute to recommendation effectiveness, we review findings from prior studies across a range of recommendation tasks. While our experiments focus on e-commerce and short-video domains, existing research has investigated the role of multimodal signals in areas such as news, food, music, and point-of-interest (POI) recommendations. These studies reveal that the value of each modality is highly task dependent and shaped by both content characteristics and the nature of user-item interactions.


In news recommendation, visual and textual modalities are often complementary, with visual features enhancing semantic richness when well-aligned with text. For instance, MM\_Rec~\cite{wu2022mm} employs the vision-language model VILBERT~\cite{lu2019vilbert}, which is powerful for its multimodal understanding capabilities, to represent news content using both visual and textual modalities. Their experiments show that incorporating both modalities improves performance compared to using either alone, suggesting that multimodal signals are particularly useful when content is diverse and expressive. The performance comparison between using image and text alone also highlights the contribution of visual information.


In contrast, food recommendation tends to benefit less from general multimodal features. HealthRec~\cite{zhang2024multi} and CLUSSL~\cite{zhang2024multi2} leverage user interaction data with modality content and ingredients information to personalize the diet recommendation. Based on the experimental result in the original paper, we compare the performance of general collaborative filtering models using only interactions with the multimodal recommendation baselines. The performance of those multimodal recommendation models is comparable or even worse than the strongest general model LightGCN, which indicates the limited contribution of modality information in the food recommendation task. Moreover, structured data like ingredients and user interactions is more informative than image or text content, which is often repetitive and less distinctive. This highlights the importance of leveraging domain-specific signals over generic multimodal inputs.


In music recommendation, the contribution of multimodal data is more nuanced. AMAE~\cite{shen2020enhancing} evaluates the contribution of modality features and the experimental results demonstrate that incorporating modality information can improve recommendation performance, but the improvement is marginal. When comparing the performance of using multimodal data and the best single modal data, the performance only improves by 0.78\%. Using a single modality instead of interaction-only data yields improvements ranging from 1\% to 1.8\%. TALKPLAY~\cite{doh2025talkplay} also examines the impact of removing different modalities, finding that playlist tokens and metadata tokens contribute most significantly to performance. Raw audio features may offer limited value and can even degrade performance when poorly represented. The authors attribute this counterintuitive finding to the use of low-level audio features. These findings reinforce the idea that the quality and semantic relevance of modality features are critical for their effectiveness.


For POI recommendation, the benefits of incorporating multimodal data remain inconclusive. MMPOI~\cite{xu2024mmpoi} selects baseline models from sequential recommendation, multimodal recommendation, and POI recommendation categories. The experimental result shows that MMPOI
excludes the multimodal feature, performs worse than the full MMPOI model, but it still achieves results that are comparable to or better than the strongest baseline. Similar observations are reported in UGC~\cite{wang2014semantic}. These experimental findings suggest that the observed performance gains are primarily driven by architectural innovations, with visual and textual content contributing little due to their weak alignment with spatial or behavioral patterns.

Overall, these studies support our conclusion that the importance of modality varies by task. Effective multimodal recommendation requires not just the inclusion of additional data sources, but thoughtful integration strategies that account for content quality, alignment, and domain-specific user behaviors.

\begin{table}[]
\caption{Size of multimodal recommendation models on \textit{Baby}. Parameter size is the number of trainable parameters.}
\label{parameter}
\centering
\begin{tabular}{lr|lr}
\toprule
Model & Parameter Size &Model &Parameter Size\\
\midrule
VBPR    & 3,226,944  & MICRO   & 33,570,752 \\
MMGCN   & 1,418,240  & BM3     & 33,570,688 \\
DualGNN & 3,742,782  & FREEDOM & 33,566,528 \\
SLMRec  & 2,028,032  & MGCN    & 33,587,392 \\
GRCN    & 4,524,478  & MGCL    & 4,471,360  \\
LightGT & 3,418,752  & MCLN    & 5,531,584  \\
LATTICE & 33,566,530 & MGCE*   & 6,960,512 \\
\bottomrule
\end{tabular}
\end{table}

\subsection{How does Model Size Influence the Multimodal Model Performance?}

Integrating multimodal information often leads to significantly larger models, as additional components are required to process and fuse text, images, and interaction data. This added complexity increases both training and inference costs, raising concerns about scalability in real-world applications. Although larger models are commonly assumed to produce better performance, our findings (Section~\ref{sec:reprod}) show that interaction-only models can perform comparable to or even better than multimodal models. This suggests that increased model size does not necessarily translate to improved effectiveness.

Moreover, by comparing Table~\ref{parameter} and Table~\ref{same setting}, we find no clear correlation between model size and performance even among multimodal models. Some smaller models outperform larger ones, like MGCE performs best with a size smaller than LATTICE, indicating that effectiveness depends more on how modality information is integrated than on the number of parameters. These observations highlight the importance of efficient, well-designed fusion strategies over simply increasing the size of model.

\begin{figure}
    \centering
    \includegraphics[width=0.8\linewidth]{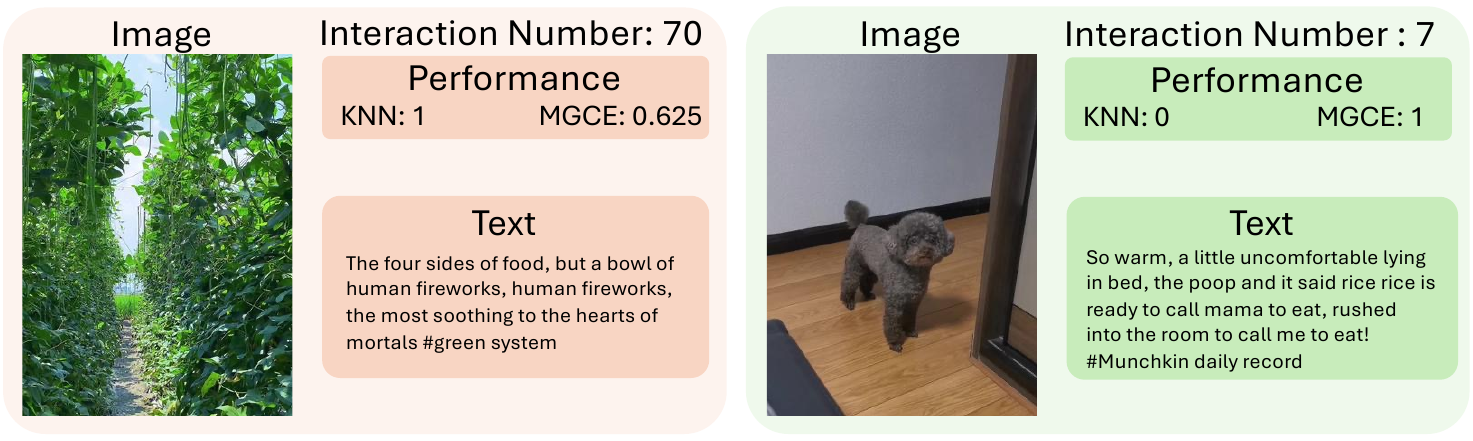}
    \caption{Example items from \textit{DY} }
    \label{dy_example}
\end{figure}

\begin{figure}
    \centering
    \includegraphics[width=0.8\linewidth]{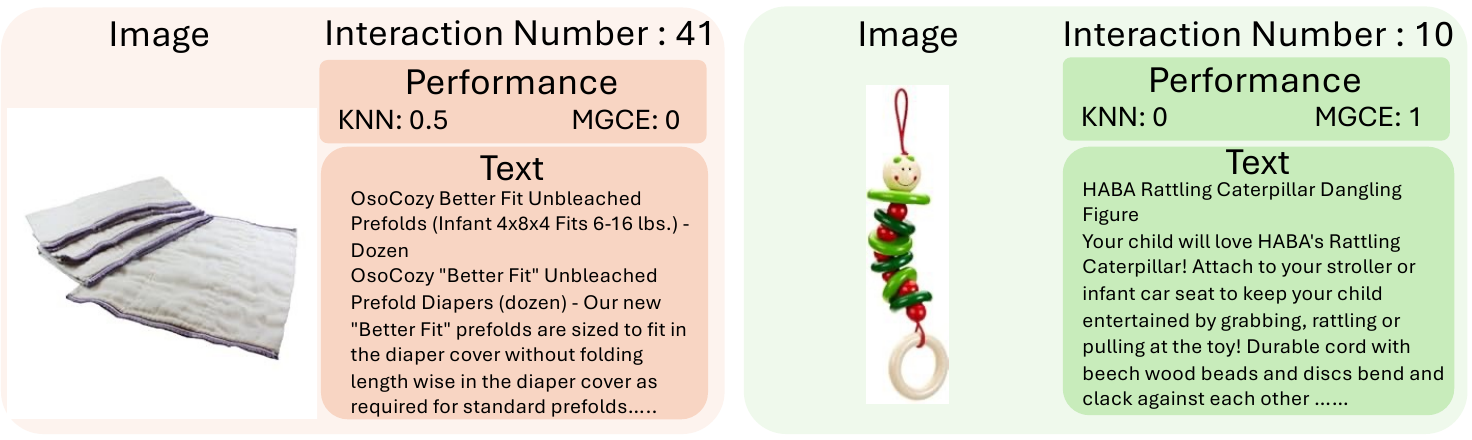}
    \caption{Example items from \textit{Baby}}
    \label{baby_example}
\end{figure}

\subsection{Case Study}
To better understand the strengths and limitations of different recommendation methods, we include case studies that offer qualitative insights into model performance. We present two representative examples from each dataset: the short-video \textit{DY} dataset (Figure~\ref{dy_example}) and the e-commerce \textit{Baby} dataset (Figure~\ref{baby_example}). In each dataset, one case illustrates a scenario where the traditional KNN model performs better (highlighted with a red background), while the other shows a case where the multimodal recommendation model yields better results (highlighted with a green background). These examples help us examine how collaborative filtering signals and multimodal features influence recommendation outcomes under different conditions. By comparing these contrasting cases, we aim to better understand when multimodal information offers a clear advantage and when simpler methods remain effective.

Our case study reveals clear patterns that help explain when traditional KNN models outperform multimodal methods, and vice versa. In cases where UserKNN performs better, the items typically have high interaction counts and low-quality multimodal content. In the \textit{DY} dataset, images often lack clear structure, and texts are vague or poorly aligned with the visuals. These weak content signals reduce the effectiveness of multimodal models, making KNN’s reliance on user behavior more reliable. A similar trend is seen in the \textit{Baby} dataset, where frequently purchased items like diapers have plain or uninformative modality features but sufficient interaction history to support effective recommendations.

In contrast, when the items have sparse interaction histories and high-quality modality data, the multimodal recommendation system MGCE outperforms KNN, where KNN struggles due to limited collaborative signals. 
In these cases, MGCE effectively leverages high-quality, well-aligned content to make accurate predictions. For example, in the \textit{DY} dataset, items with appealing visuals such as a cute dog, along with matching descriptive text, attract user interest despite minimal interaction data. Similarly, in the \textit{Baby} dataset, toys with engaging images and detailed descriptions benefit from MGCE’s ability to extract meaningful semantic cues. These cases demonstrate that multimodal models are especially advantageous when interaction data is limited. 

These findings suggest that multimodal models are more effective when content features are rich and coherent, and interaction data is sparse. This aligns with prior studies showing that user behavior reflects historical preferences, while modality information can influence interaction decisions by offering additional context and emotional appeal~\cite{li2014impact,sun2024impact}.



\section{Future Directions of Multimodal Recommendation}
The era of information explosion on online platforms, combined with the growing utility of multimodal data, has accelerated the development of multimodal recommendation systems. Multimodal information has shown promise in addressing key challenges such as the cold start and data sparsity problems. Based on our prior review and analysis, we identify several challenges and outline potential future research directions in the field of multimodal recommendation systems.

\subsection{Modality Fusion}
The effectiveness of incorporating multimodal information to enhance recommendation performance has been well demonstrated in prior studies~\cite{he2016vbpr,ong2025spectrum}. Multimodal data typically encompasses diverse aspects of data and contains complementary information about the items. Recent approaches~\cite{liu2024alignrec,guo2024lgmrec} typically employ explicit alignment mechanisms to facilitate the fusion of multimodal data, aiming to refine user and item embeddings. However, we argue that the contribution and relevance of each modality can vary significantly across different tasks. Experimental results indicate that low-quality modality features or suboptimal fusion strategies can lead to a decline in recommendation performance. 

Furthermore, the practical deployment of multimodal recommendation systems requires robustness, particularly in scenarios where data from certain modalities is missing or noisy for specific items. For instance, MILK~\cite{bai2024multimodality} introduces an alignment module that leverages multimodal invariant learning on pretrained multimedia features to ensure semantic consistency across modalities. Similarly, SiBraR~\cite{ganhor2024multimodal} adopts a single-branch network architecture with shared weights across modalities to mitigate the effects of missing or noisy data. Another line of research employs generative modules to enhance modality-specific representations under conditions of incomplete or noisy input~\cite{wang2018lrmm, lin2023contrastive, yi2021multi}. Alternatively, some studies explore imputation techniques to reconstruct missing features prior to generating recommendations~\cite{wang2018lrmm, malitesta2024we}.

This introduces a key challenge: how to effectively fuse multimodal data to capture complementary information from heterogeneous sources, while mitigating the influence of noise or modality missing. For future work, research should focus on modality selection mechanisms, where researchers could select and assign weights to modalities on a test dataset before applying it to the whole dataset, based on the specific recommendation scenario and context. Moreover, instead of relying on manual experimentation to determine useful modalities, we could integrate adaptive weighting strategies or attention mechanisms within the model architecture to automatically filter and prioritize modality information. Such methods would enhance the robustness and generalizability of multimodal recommendation systems across diverse tasks and data conditions.

\subsection{Integrate Collaborative and Modality Signals}

Our empirical results reveal that simply combining collaborative and modality signals is insufficient to guarantee consistent performance improvements. MGCL~\cite{liu2023multimodal} also indicates that merging collaborative signals and modality-specific preference as a whole to perform message passing between nodes will introduce noise in the modality feature to collaborative signals. The effectiveness of integration strategies (e.g., early fusion vs. ensemble learning) differs across models, and poor fusion can even degrade performance. These findings suggest that the modality signals and collaborative signals are not always aligned or equally useful throughout the training process. We need to find a way that integrate collaborative signal and modality signals efficiently and not influence the basic performance of collaborative signals. 

To address this, we propose exploring curriculum learning paradigms for multimodal recommendation. Instead of fusing all signals from the outset, curriculum learning can guide the model to gradually introduce modalities during training based on their informativeness and reliability. This staged integration may help prevent early overfitting to noisy or redundant modalities and better align modality representations with collaborative preferences. Future research can investigate how to design modality difficulty metrics, dynamic fusion schedules, and curriculum-aware architectures to optimize the integration process in an adaptive manner.

\subsection{User Modality Preference}
Intuitively, users tend to pay varying degrees of attention to different modalities when making decisions. Some are more influenced by visual content, while others rely heavily on textual descriptions. However, most current multimodal recommendation models overlook this diversity and adopt a uniform fusion strategy across all users, assuming equal importance of each modality, which can lead to suboptimal representations when user preferences are modality-specific. To address this limitation, several recent models have introduced mechanisms to dynamically adjust the influence of each modality based on individual user profiles. For example, MAML~\cite{liu2019user2} utilizes an attention neural network to model users' diverse preferences on different aspects of various items. The Hierarchical User Intent Graph Network~\cite{wei2021hierarchical} exhibits user intents in a hierarchical graph structure to profile personal interests. Future research can explore user-aware attention networks or adaptive gating modules to dynamically adjust the influence of each modality based on individual user profiles. The system can tailor recommendation outputs to align with each user's preferred modality focus, thereby enhancing both accuracy and personalization. 

\subsection{Stage-Specific Multimodal Architecture Design}
Experimental results reveal that multimodal features are more beneficial during the recall stage than in the reranking stage, where their effect often diminishes or even introduces noise. These findings highlight the potential for developing stage-specific models that better leverage multimodal information to rerank the candidates. ARMMT~\cite{xu2024advancing} employs an attention-based multimodal fusion technique and an auxiliary ranking-aligned task to enhance item representation and improve targeting capabilities. MM-R5~\cite{xu2025mm} proposes a task-specific reward framework, which includes a reranking reward designed for multimodal candidates and a composite template-based reward to further improve reasoning quality. Future research can develop stage-specific multimodal modeling, where multimodal information is used to optimize the recall stage, while the reranking stage is refined through knowledge-distilled architectures or tailored reward functions. Such separation may reduce model complexity and mitigate overfitting in the reranking phase.

\subsection{Cross Domain Multimodal Recommendation}

Current models are often tailored to a single domain. Cross-domain recommendation systems aim to enhance performance in a target domain by leveraging user behavior or content information from other domains or platforms~\cite{liang2022hierarchical}. Traditional approaches typically rely on explicit overlap between domains, such as shared users or items, to enable knowledge transfer. However, these methods are limited in scalability and applicability when such overlap is sparse or unavailable. Recent advances have moved toward learning universal representations for users and items that are transferable across domains. For example, UniSRec~\cite{hou2022towards} demonstrates that universal item embeddings can be learned from textual information alone, allowing for effective recommendations even when no common users or items exist between source and target domains.

Looking forward, multimodal information like images, text, and audio offers complementary signals that could significantly enhance the quality of these universal representations. Integrating multimodal content into cross-domain learning opens new possibilities for aligning semantically similar items across domains. For instance, an item in one platform and a similar product in another could be matched through shared visual or textual features, enabling cold-start transfer without user overlap. Future research can explore the cross-domain transferability of multimodal representations to optimize the recommender for a new domain. This approach promises to broaden the applicability and effectiveness of cross-domain recommendation systems, harnessing a richer variety of data sources to provide more accurate recommendations in new domains.

\subsection{Research and Application Gap}

Despite notable progress in multimodal recommendation research, a noticeable gap remains between academic advancements and real-world industry deployment. One primary challenge concerns the accessibility and quality of datasets. Academic studies often rely on publicly available datasets such as Amazon, which contain low-quality images, align with experimental results that show greater performance gains from textual modalities than from visual ones. In contrast, industry can access real-world high-quality datasets that allow models to more accurately capture user preferences. However, real-world applications must also deal with missing modalities and noisy data, which further complicate deployment. Another major barrier is computational cost. Many multimodal recommendation models depend on large-scale pretrained encoders, complex fusion mechanisms, or graph neural networks, all of which are difficult to scale within practical constraints on latency, memory, and energy consumption. Future research should aim to bridge this gap by developing robust, scalable models that strike a balance between performance and efficiency. Improving the computational efficiency of processing high-dimensional tensors and multimodal inputs is essential for advancing multimodal recommendation systems from academic prototypes to real-world applications.

\section{Conclusion}
In conclusion, this study provides a comprehensive evaluation of multimodal recommender systems, revealing that while multimodal data can enhance recommendation performance, the improvements are often modest and context-dependent. We find that multimodal models are particularly effective in sparse interaction scenarios and in the recall stage of the recommendation pipeline, where content signals offer greater value. The importance of each modality varies by task, with text features proving more beneficial in e-commerce settings and image features playing a larger role in short-video domains. Ensemble-Based Learning consistently outperforms Fusion-Based approaches, emphasizing the value of maintaining modality-specific signals before integration. Additionally, our analysis shows that larger model size does not guarantee better performance, highlighting the need for efficient design over complexity. A broader review across domains such as news, food, music, and POI, along with case studies, further supports these findings and underscores the importance of task-aware, well-integrated multimodal strategies.

Despite the breadth of our analysis, several limitations remain that may affect the generalizability and completeness of our findings. First, variations in dataset splits, evaluation metrics, and experimental protocols across studies can introduce inconsistencies, making direct comparisons challenging. Second, the quality, granularity, and alignment of multimodal data (e.g., noisy text, low-resolution images, or misaligned modalities) can significantly influence model performance and may obscure the true potential of multimodal integration. Third, our evaluation focuses primarily on offline performance metrics, which may not fully capture real-world effectiveness, including aspects like user satisfaction, interpretability, or fairness.

To address these challenges, future research should consider adopting unified benchmarking protocols, leveraging a more diverse set of datasets across domains, and incorporating richer, higher-quality modality data. Additionally, exploring online evaluation settings and user studies would provide more realistic insights into system impact. Advancing modality-aware pretraining, adaptive fusion mechanisms, and lightweight architectures may also improve performance, scalability, and interpretability in practical deployments. Despite these limitations, we believe our work lays a solid and practical foundation for advancing the development of effective, efficient, and domain-adaptable multimodal recommendation systems.


\bibliographystyle{ACM-Reference-Format}
\bibliography{sample-base}


\end{document}